\documentclass[11pt]{article}

\usepackage{amsmath,amssymb}
\usepackage{array}
\usepackage{graphicx}
\usepackage{graphicx}
\usepackage{wrapfig}
\usepackage{color}
\usepackage{tensor}
\usepackage{hyperref}
\usepackage[sort,compress]{cite}
\usepackage{tikz}
\usepackage[T1]{fontenc}
\usetikzlibrary{decorations.pathmorphing, decorations.pathreplacing, decorations.shapes} 

\numberwithin{equation}{section}

\renewcommand{\d}{\mathrm{d}}

\newcommand{\region}{$\{x^+>0,x^-<0\}$\ }

\begin{document}
\title{\textbf{
Hybrid quantum states in 2d dilaton gravity}}
\vspace{0.5cm}
\author{ \textbf{ Yohan Potaux$^{a}$, Debajyoti Sarkar$^{b}$ and Sergey N. Solodukhin$^{a,c}$ }} 

\date{}
\maketitle
\begin{center}
    \emph{$^{a}$Institut Denis Poisson UMR 7013,
  Universit\'e de Tours,}\\
  \emph{Parc de Grandmont, 37200 Tours, France} \\
\vspace{0.2cm}
\emph{$^{b}$Department of Physics}\\
  \emph{Indian Institute of Technology Indore}\\
  \emph{Khandwa Road 453552 Indore, India}\\
  \vspace{0.2cm}
  \emph{$^{c}$Institute for Theoretical and Mathematical Physics, } \\
\emph{Lomonosov Moscow State University, 119991 Moscow, Russia}
\end{center}

\vspace{0.5cm}
  \texttt{ yohan.potaux@lmpt.univ-tours.fr, dsarkar@iiti.ac.in, solodukh@lmpt.univ-tours.fr}

\begin{abstract}

The classical black hole spacetime is modified semiclassically, depending strongly on the choice of the quantum states. In particular, for the Boulware state the spacetime often takes a wormhole structure mimicking closely a spacetime with a horizon. In this paper, in the context of the two-dimensional dilaton RST model, we consider all possible important interplays between the Hartle-Hawking, Unruh and Boulware quantum states. Special attention is given to the hybrid states made up of quantum fields either in the Hartle-Hawking or Unruh states, and some non-physical fields (with the wrong sign in the kinetic term in the action) in the Boulware state. We present a detailed analysis of the semiclassical geometry in all these cases paying attention to the presence or absence of horizons, curvature singularities and to the geodesic completeness of the spacetime. In the space of parameters specifying the generic quantum state, we find a wide domain (with dominating non-physical fields) where the semiclassical geometry represents a geodesically complete, asymptotically flat causal diamond, free of horizon or curvature singularity. However, a distant observer still finds Hawking radiation at asymptotic infinity. In the Unruh-Boulware hybrid state solution, we find that the energy flux at asymptotic infinity receives important corrections from its thermal behavior, leading to information recovery as we go from early to late retarded times. As a result, the corresponding entropy shows a typical Page curve bahavior.

\end{abstract}

\newpage
\tableofcontents
\pagebreak

\section{Introduction}
The information paradox, or problem of information loss in black holes, presents a conflicting point between the previsions of general relativity and the laws of quantum mechanics. Historically, the information loss problem was posed in the framework of semiclassical gravity where the properties of quantum fields are studied on a classical background geometry, and this is the framework where the information problem is most concretely posed and has been investigated the most. Even though the problem will be fully solved only within a set-up of quantum gravity, one can still wonder whether the change in the classical geometry due to the back-reacting quantum fields may have something illuminating to say about the information paradox. A first order thought would seem to suggest that it is not possible, not to mention the fact that such an approach is perturbative by-construction. However, such conclusions turn out to be naive, and in fact, with certain conditions on lower dimensional systems (and also in some higher dimensional cases), the back-reacted spacetime solutions can be understood as non-perturbative and complete. In two dimensions, string inspired classical theories \cite{Mandal:1991tz,Witten:1991yr} were indeed a subject of collaborative focus in the early to mid-nineties which led to the study of quantum fields in their background (for a review, see e.g. \cite{Grumiller:2002nm}). Various studies of two-dimensional gravity systems have once again resurfaced in the past few years (in the context of black hole information paradox), and our present work can be understood as an entry to this vast and unfolding topic. In particular, we study a two-dimensional model of semi-classical gravity, where quantum fields propagate on a classical spacetime, thereby influencing its geometry in a manner that reinstates unitary evolution of the black hole system. We have already studied some aspects of this model in \cite{potaux_quantum_2022}, focusing on static solutions, and here we will review these static solutions with a different coordinate system, which will allow us to discuss dynamical situations. This paper was written in parallel to a shorter one \cite{potaux_space-time_2022} which focuses on one of the most interesting scenarios regarding the information paradox.

The model we chose to study is the semi-classical RST model \cite{russo_endpoint_1992}, which extends the classical CGHS model \cite{callan_evanescent_1992} by taking into account conformal anomalies. The way to do this is unclear in four dimensions but in two dimensions it is done by adding the  Polyakov action \cite{Polyakov:1981rd} to the CGHS one. This additional term is non-local and this translates into some unfixed parameters when solving the gravitational equations, which can be determined by choosing a state for the quantum matter. The states we have already discussed in previous publications are 

\medskip

\noindent{\it The Hartle-Hawking state}:  it contains  thermal radiation at infinity and the stress-energy  tensor is regular at the horizon. It describes a black hole in thermal equilibrium with the Hawking radiation.
 
\vspace{15pt}
  
\noindent {\it The Boulware state}: the stress-energy tensor is  vanishing at infinity and there is no radiation there. However, considered on a classical black hole metric, it is singular at the horizon.

\vspace{15pt}
 
\noindent  In the present paper we will also focus on 
  
\vspace{15pt}

\noindent {\it The Unruh state}: the stress-energy tensor is regular only at the future horizon; it vanishes at past null infinity and there is a thermal flux of radiation at future null infinity. The Unruh (U) state is usually considered to describe the process of the black hole evaporation.

\medskip

For physical quantum particles in equilibrium with the classical black hole, the Hartle-Hawking (HH) state is the most natural and suited one, as an observer at infinity should detect the radiation due to these particles. The Boulware state (B) is the most appropriate one when considering non-physical particles, such are ghosts, in the spacetime. By ghosts in the present paper we mean the particles whose kinetic terms has a negative sign. Indeed an observer at infinity should not detect these particles and the Boulware state is the only one with this property. The problem of ghosts in black hole evaporation was discussed in e.g. \cite{strominger_fadeev-popov_1992,bilal_liouville_1993,Liberati:1994za}. As suggested briefly in our previous paper, it is possible to consider a \emph{hybrid} quantum state where both physical and non-physical particles are present, each in its own quantum state (HH or U for physical particles and B for non-physical ones). In this paper we analyze what new ideas these hybrid states can bring to the discussion of the information paradox.

Our aim here is to consider all the combinations and complexities of these states. The usual approach, where a classical black hole background and the Hawking radiation are taken separately, is obviously fundamentally incomplete. That is why our primary goal is to address the issue of the back-reaction of the Hawking radiation on the spacetime as a part of the self-consistent picture.
The first principal question here is whether the black hole horizon is still present in the back-reacted geometry. Previous studies have shown that the answer to this question
is sensitive to the choice of the vacuum state for the quantum fields in question. In the case of the Hartle-Hawking state the horizon is still there rather similarly to the
classical spacetime, although the parameters of the spacetime geometry receive certain corrections and the geometry is deviated from the classical one (even leading to naked singularities in some case).
On the other hand, the situation is radically different when the Boulware state is chosen. Our analysis here makes it fairly clear that the classical horizon is no more present in this case.
Indeed, the semiclassical geometry depends on the parameters present in the theory and quite generally, the following possibility is realized: a timelike or a lightlike singularity  or
a narrow throat is formed. The narrow throat represents a  black hole mimicker, a horizon-less spacetime 
that models rather closely the general properties of a black hole.  

The main focus of the present paper is on the hybrid quantum state. We identify a situation in which the
back-reacted spacetime is a complete diamond similar to the Minkowski spacetime without an horizon nor a singularity. 
No information can be fundamentally lost in such a spacetime. However, looking from a distance this spacetime behaves like a black hole during a relatively long 
retarded time. A distant observer does detect a thermal radiation similar to the Hawking radiation. So that the usual information problem would seem to arise for the observer during this period of time.
However at later retarded times important deviations from thermality start to show up in the asymptotic radiation and effectively leads to restoration of the information 
and to the expected form of the Page curve for the asymptotic radiation. When the hybrid state consists of particles from the HH and B states, this is the picture presented in  \cite{potaux_space-time_2022}. In the present paper, we have filled-in certain technical details which we omitted in  \cite{potaux_space-time_2022}. Afterwards, we generalize the discussion to the hybrid Unruh-Boulware state that has more realistic properties compared to the Hartle-Hawking counterpart,
and represents a time dependent, evolving situation.  This investigation is new and it represents our main result in this paper. We expect that the two-dimensional  case discussed here is indicative of what may happen in a more realistic four-dimensional
situation, that is less technically accessible (although we refer to  \cite{Berthiere:2017tms} as a step towards a realisation of the present program in four dimensions).

A brief outline of our present work is the following. We start with a discussion of the (semi)classical CGHS (and RST) models in sections \ref{Section_ClassicalModel} and \ref{sec:RST}, which illustrates how the relevance of the quantum state arises when going from the classical to the the semiclassical analysis. Afterwards, in section \ref{sec:singlestate}, we start by reviewing the static and dynamical solutions for the classical CGHS model and the RST model with only one type of particles in a single quantum state (namely HH or B). The dynamical situations are obtained by sending in a shock wave of classical matter into the static spacetimes, and this analysis is done quite easily in the conformal gauge. In sections \ref{SectionHybridRST} and \ref{sec:hybridunruh}, we then investigate how considering a hybrid quantum state modifies the solutions. There are three possible cases to consider, according to the kind of particles (physical or non-physical) that predominates the back-reaction on the spacetime. As it turns out, the most interesting situation is when the number of non-physical particles exceeds the number of physical particles, which is why it was the focus of our shorter paper \cite{potaux_space-time_2022}. Notably, there we explained that the static spacetime is completely regular with a geodesically complete, wormhole type structure, while in the dynamical case an apparent horizon appears. The energy balance in asymptotic region is also interesting, and has been studied for all cases in section \ref{sec:energy}. For the hybrid case, if the spacetime is static, a thermal radiation is detected at infinity; while for the dynamical situation, the radiation deviates from thermality as time passes. In this case the change in entropy of radiation exhibits the behavior expected from the Page curve \cite{page_information_1993,page_time_2013}. The study of the Page curves for black holes and the importance of wormhole type structures has been a topic of recent investigation for both two and higher dimensions, fueled by the new developments in the direction of quantum extremal surfaces (see for example the reviews \cite{Almheiri:2020cfm,Kundu:2021nwp}. And for an application towards black hole firewall \cite{Almheiri:2012rt,Almheiri:2013hfa}, see\cite{Germani:2015tda}). The references above are of course far from complete, but for some concluding remarks along this line, with an enumeration of our results in various cases, see section \ref{sec:conclude}.

\section{The classical model}
\label{Section_ClassicalModel}

We begin our discussion with the classical CGHS model \cite{callan_evanescent_1992} whose action is 
\begin{equation}
 \mathcal{S}_0 = \frac{1}{2\pi}\int_M \mathrm{d^2}x\, \sqrt{-g} \bigg\{e^{-2\phi}\big[R + 4(\nabla\phi)^2 + 4\lambda^2\big] - \frac{1}{2}(\nabla f)^2 \bigg\}
 \,,
 \label{CGHS_Action}
\end{equation}
where $\phi$ is the dilaton scalar field, $f$ the matter field and $\lambda^2$ a cosmological constant. This model has origins in string theory but is interesting in its own right as a toy model of two-dimensional gravity. Note that the coupling strength is given by $e^{\phi}$.

As noted in \cite{russo_black_1992} it is instructive to compare this model with the action obtained by considering a spherically symmetric metric in four-dimensional gravity
\begin{equation}
 \bar{g} = g_{\mu\nu}\d x^\mu \d x^\nu + r^2(x^0,x^1)(\d\theta^2 + \sin^2\theta\d\varphi^2)
 \,, \quad
 \mu,\nu = 0,1
 \,.
 \label{4D_Spherical_Metric}
\end{equation}
If we compute the Einstein-Hilbert action for this metric and define $\phi = -\ln(\lambda r)$ we get
\begin{equation}
 \mathcal{S}_{EH} = \frac{1}{2\kappa}\int\d^4x\,\sqrt{-\bar{g}}\bar{R}
 = \frac{1}{2\pi}\int\d^2x\,\sqrt{-g}e^{-2\phi}\bigl[R + 2(\nabla\phi)^2 + 2\lambda^2e^{2\phi}\bigr]
 \,, \quad
 \kappa = \biggr(\frac{2\pi}{\lambda}\biggr)^2
 \,,
\end{equation}
which is similar to \eqref{CGHS_Action}, therefore one can think of $\frac{e^{-\phi}}{\lambda}$ as the radius of a transverse sphere at $(x^0,x^1)$.
See also  \cite{Kazakov:1993ha} for more discussions on this effective 2d model of gravity.

This analogy can help us to find a condition for the existence of apparent horizons in terms of the dilaton field $\phi$. Indeed an apparent horizon is defined as the outer boundary of a region of trapped surfaces, \textit{i.e.}\ surfaces whose area decreases along any future direction. Using the four-dimensional metric \eqref{4D_Spherical_Metric} we see that the area of a transverse sphere at $(x^0,x^1)$ is $\frac{4\pi}{\lambda^2}e^{-2\phi}$. In null coordinates $x^\pm = x^0\pm x^1$ this transverse sphere is trapped if its area is decreasing in both null directions, \textit{i.e.}\ if $\partial_\pm \phi > 0$. Going back to the two-dimensional theory while using the conformal gauge (see Appendix \ref{AppendixConformalCoord}) where the metric takes the form
\begin{equation}
 \mathrm{d}s^2 = -e^{2\rho}\mathrm{d}x^+\mathrm{d}x^-
 \,,
\end{equation}
there will be a region of trapped points where $\partial_\pm\phi >0$.
Since $(\nabla\phi)^2 = -4e^{-2\rho}\partial_+\phi\partial_-\phi$ this implies that $(\nabla\phi)^2 < 0$, which means that the vector field $\nabla\phi$ is timelike in this region. The corresponding apparent horizon is the boundary of this region, where therefore $(\nabla\phi)^2 = 0$.

However this condition is not enough to define an apparent horizon as one also has to check that $(\nabla\phi)^2$ changes sign when crossing this boundary, it is positive outside the horizon and negative inside. If $(\nabla\phi)^2$ vanishes on an hypersurface but is positive on both sides then there is no apparent horizon but what we will call a Type I wormhole: the radius of the transverse sphere has a minimum there but it increases on both sides. To summarize we have the following definitions:

\medskip

\noindent \textit{Apparent horizon}: hypersurface on which $(\nabla\phi)^2 = 0$ with $(\nabla\phi)^2 > 0$ on one side (the outside of the horizon) and $(\nabla\phi)^2<0$ on the other (the inside).

\vspace{15pt}

\noindent \textit{Type I wormhole}: region containing an hypersurface where $(\nabla\phi)^2=0$ with $(\nabla\phi)^2 > 0$ on both sides. This represents a local minimum of the dilaton function $\phi(x^+, x^-)$. Taking the analogy between the dilaton $\phi$ and the radial coordinate $r$ in the 4d picture, this definition corresponds to a usual understanding of the wormhole.

\vspace{15pt} 

\noindent Finally we define a 

\vspace{5pt} 

\noindent \textit{Type II wormhole} following \cite{damour_wormholes_2007}	as follows: if the spacetime has a timelike Killing vector field $\xi$, which we will normalize by imposing that $\xi = \partial_t$ in flat asymptotic coordinates $(t,x)$, and the $(tt)$ metric function $g = -\xi^2$ has a minimum in some region, then this region corresponds to the throat of a wormhole, where time flows slowly with respect to the time of an external observer. If this minimum of $g$ is exponentially small and the spacetime is asymptotically flat on both sides (which in particular means it has to be singularity free) then this is a black hole mimicker, as discussed in \cite{damour_wormholes_2007}. Note that this definition can be used in static case and is only approximative 
in a dynamical situation.

We note that this discussion is motivated by the earlier study in four dimensions where the black hole mimicker is represented by a spacetime in which wormhole structures of both type I and II
are present at the same spherical surface. This is however not a necessary condition  to have a good black hole mimicker as the examples discussed in the present paper show.
So that a wider class of mimickers is possible to exist.

\medskip

Now we will review some results discusses in the original CGHS article \cite{callan_evanescent_1992}.
\subsection{Equations of motion}
As we work in two dimensions, the Einstein tensor is identically zero so the equations of motion for the metric are given by the vanishing of the stress-energy tensor, \textit{i.e.}
\begin{align}
 T^{(0)}_{\mu\nu} & \equiv - \frac{2}{\sqrt{-g}}\frac{\delta \mathcal{S}_0}{\delta g^{\mu\nu}}
 \nonumber
 \\
 & = \frac{2}{\pi}e^{-2\phi}\Big\{-\nabla_\mu\nabla_\nu\phi + g_{\mu\nu}\big[\square\phi - (\nabla\phi)^2 + \lambda^2\big]\Big\}
 - \frac{1}{4\pi}g_{\mu\nu}(\nabla f)^2
 + \frac{1}{2\pi}\partial_\mu f \partial_\nu f
 \nonumber
 \\
 & = 0
 \,.
 \label{CGHS_StressEnergy0}
\end{align}
Note that we use the same normalization of the stress-energy tensor as in \cite{potaux_quantum_2022}. There is also the equation of motion for the dilaton, given by
\begin{equation}
 R + 4\big\{\square\phi - (\nabla\phi)^2 + \lambda^2\big\} = 0
 \,,
 \label{CGHS_DilatonEquation}
\end{equation}
and the one for the matter
\begin{equation}
 \square f = 0
 \,.
 \label{CGHS_MatterEquation}
\end{equation}
The trace of the stress-energy tensor is
\begin{equation}
 g^{\mu\nu}T^{(0)}_{\mu\nu} =
 \frac{2}{\pi}e^{-2\phi}\Big\{\square\phi - 2(\nabla\phi)^2 + 2\lambda^2\Big\} = 0
 \,,
\end{equation}
and combining this with the dilaton equation \eqref{CGHS_DilatonEquation} leads to
\begin{equation}
 R + 2\square\phi = 0
 \,.
 \label{CGHS_Trace}
\end{equation}

\subsection{Killing vector}
Let us define the vector field $\xi$ by
\begin{equation}
 \lambda\xi_\mu \equiv \tensor{\epsilon}{_\mu^\nu}\nabla_\nu\phi
 \,,
 \label{CGHS_Killing_Definition}
\end{equation}
where $\epsilon$ is the Levi-Civita tensor defined by $\epsilon_{01} = \sqrt{-g}$ and $\epsilon_{\mu\nu} = -\epsilon_{\nu\mu}$. 
The normalization in (\ref{CGHS_Killing_Definition}) is chosen so that $\xi=\partial_t$ asymptotically. 
Using the equations of motion for the metric \eqref{CGHS_StressEnergy0} and the dilaton \eqref{CGHS_DilatonEquation} we obtain that
\begin{equation}
 \nabla_\mu\nabla_\nu\phi =
 - \frac{1}{4}Rg_{\mu\nu} - \frac{e^{2\phi}}{2}\bigg\{\frac{1}{4}g_{\mu\nu}(\nabla f)^2 - \frac{1}{2}\partial_\mu f\partial_\nu f\bigg\}
 \,,
\end{equation}
so
\begin{align}
 \lambda(\nabla_\mu\xi_\nu + \nabla_\nu\xi_\mu) & =
 \tensor{\epsilon}{_\nu^\rho}\nabla_\mu\nabla_\rho\phi
 + \tensor{\epsilon}{_\mu^\rho}\nabla_\nu\nabla_\rho\phi
 \nonumber
 \\
 & = -\frac{1}{4}\bigg\{R + \frac{e^{2\phi}}{2}(\nabla f)^2\bigg\}
 \underbrace{(\epsilon_{\nu\mu} + \epsilon_{\mu\nu})}_{=0}
 + \frac{e^{2\phi}}{4}(\tensor{\epsilon}{_\nu^\rho}\partial_\mu f + \tensor{\epsilon}{_\mu^\rho}\partial_\nu f)\partial_\rho f
 \nonumber
 \\
 & = \frac{e^{2\phi}}{4}(\tensor{\epsilon}{_\nu^\rho}\partial_\mu f + \tensor{\epsilon}{_\mu^\rho}\partial_\nu f)\partial_\rho f
 \,.
\end{align}
Therefore if there is no classical matter ($f=0$) $\xi$ is a Killing vector and its norm is given by
\begin{equation}
 (\nabla\xi)^2 = -\frac{1}{\lambda^2}(\nabla\phi)^2
 \,.
\end{equation}
The vanishing of $\xi^2$ will indicate the presence of a Killing horizon, in this classical model it corresponds to the vanishing of $(\nabla\phi)^2$ so that it coincides with the apparent horizon condition defined previously, provided that the norm of $\xi$ changes sign when crossing this horizon.
\subsection{Conformal gauge}
Since we work in two dimensions we can put the metric under the conformal form (see Appendix \ref{AppendixConformalCoord})
\begin{equation}
 \d s^2 = -e^{2\rho}\d x^+ \d x^-
 \,,
\end{equation}
and in these coordinates the equations of motion become
\begin{equation}
 \left\{
 \begin{aligned}
  T^{(0)}_{\pm\pm} & = \frac{1}{\pi}e^{-2\phi}(4\partial_\pm\rho \partial_\pm\phi -2\partial_\pm^2\phi )
  + \frac{1}{2\pi}(\partial_\pm f)^2 = 0
  \,, \\
  T^{(0)}_{+-} & = \frac{1}{\pi}e^{-2\phi}(2\partial_+\partial_-\phi - 4\partial_+\phi\partial_-\phi - \lambda^2e^{2\rho}) = 0
  \,,
 \end{aligned}
 \right.
 \label{CGHS_ConformalStressEnergy}
\end{equation}
while the dilaton and matter equations respectively become
\begin{equation}
 -4\partial_+\partial_-\phi + 4\partial_+\phi\partial_-\phi + 2\partial_+\partial_-\rho + \lambda^2e^{2\rho} = 0
 \,,
\end{equation}
and
\begin{equation}
 \partial_+\partial_- f = 0
 \,.
\end{equation}
Finally the trace condition \eqref{CGHS_Trace} gives
\begin{equation}
 \partial_+\partial_- (\rho-\phi) = 0
 \,,
\end{equation}
and this allows us to write $2\rho = 2\phi + j$ where $j$ satisfies $\partial_+\partial_-j = 0$, which means that it is takes the form
\begin{equation}
 j = j_+(x^+) + j_-(x_-)
 \,.
\end{equation}
By changing the coordinates to
\begin{equation}
 \tilde{x}^\pm (x^\pm) = \int^{x^\pm} e^{j_\pm(u)}\d u
 \,,
\end{equation}
we get
\begin{equation}
 \d s^2 = -e^{2\phi}\d \tilde{x}^+\d \tilde{x}^-
 \,,
\end{equation}
which means that we can set $j = 0$ and $\rho = \phi$.

In this new system of coordinates the $T^{(0)}_{+-}$ equation becomes
\begin{equation}
 e^{-2\phi}(2\partial_+\partial_-\phi - 4\partial_+\phi\partial_-\phi) = \lambda^2
 \,,
\end{equation}
\textit{i.e.}
\begin{equation}
 \partial_+\partial_- e^{-2\phi} = -\lambda^2
 \,,
\end{equation}
and we can integrate this equation two times with respect to $x^+$ and $x^-$ to get that
\begin{equation}
 e^{-2\phi} = u - \lambda^2 x^+x^-
 \,,
\end{equation}
where
\begin{equation}
 u = u_+(x^+) + u_-(x^-)
 \,.
\end{equation}
The matter equation is immediately integrated to get that
\begin{equation}
 f = f_+(x^+) + f_-(x^-)
 \,.
\end{equation}
Now we can insert all this into the equations $T^{(0)}_{\pm\pm} = 0$. This gives, for $T^{(0)}_{++}$,
\begin{equation}
 e^{-2\phi}\big\{4(\partial_+\phi)^2 - 2\partial_+^2\phi\big\} + \frac{1}{2}(\partial_+f)^2 = 0
 \,,
\end{equation}
\textit{i.e.}
\begin{equation}
 \partial_+^2(e^{-2\phi}) + \frac{1}{2}(\partial_+f)^2 = 0
 \,,
\end{equation}
so that $u_\pm$ has to satisfy
\begin{equation}
 u_\pm'' = - \frac{1}{2}(\partial_\pm f)^2
 \,.
 \label{CGHS_u+-Equation}
\end{equation}
\subsection{Solution without classical matter}
If there is no matter ($f=0$) then $u^\pm$ is a linear function of $x^\pm$ so
\begin{equation}
 u = \lambda^2(c_+x^+ + c_-x^-) + C
 \,, \quad
 c_\pm,\, C \in \mathbb{R}
 \,,
\end{equation}
and
\begin{equation}
 e^{-2\phi} = C -\lambda^2(x^+x^- - c_+x^+ - c_-x^-)
 \,.
\end{equation}
We can then translate the coordinates by defining $\tilde{x}^\pm \equiv x^\pm - c_\mp$ and set $C + \lambda^2c_+c_- \equiv \frac{M}{\lambda}$ to finally obtain
\begin{equation}
 e^{-2\phi} = \frac{M}{\lambda} - \lambda^2x^+x^-
 \,.
 \label{CGHS_OmegaSolution}
\end{equation}
Therefore the metric is given by
\begin{equation}
 \d s^2 = - \frac{\d x^+\d x^-}{\frac{M}{\lambda} - \lambda^2x^+x^-}
 \,, \quad
 M \geq 0
 \,,
 \label{CGHSConformalStaticSolution}
\end{equation}
which is the well-known solution for the CGHS model. The parameter $M$ corresponds to the ADM mass of the spacetime so we take $M \geq 0$. Note that $e^{-2\phi} > 0$ so that the spacetime corresponds at most to the region $ x^+x^- < \frac{M}{\lambda^3}$. We can compute the components of the Killing vector field $\xi$ defined by \eqref{CGHS_Killing_Definition} and we get
\begin{equation}
 \xi = \lambda(x^+\partial_+ - x^-\partial_-)
 \,.
 \label{CGHS_Killing_Coordinates}
\end{equation}

\subsection{Minkowski spacetime}
\label{Section_Minkowski_CGHS}
Let us check that taking $M=0$ in \eqref{CGHS_OmegaSolution} gives the Minkowski spacetime. In this case we have $e^{-2\phi} = -\lambda^2x^+x^- > 0$, so that the spacetime corresponds to the region $ x^+x^- < 0$, and the metric is
\begin{equation}
 \d s^2 = \frac{\d x^+\d x^-}{\lambda^2x^+x^-}
 \,.
 \label{CGHS_Minkowski_Metric}
\end{equation}
There are two distinct regions where $x^+x^- < 0$, namely \region and $\{x^+ <0, x^->0\}$ which are causally disconnected and separated by a coordinate singularity at $(x^+,x^-)=(0,0)$. These two regions are the same as $x^+$ and $x^-$ are symmetric in the solution \eqref{CGHS_Minkowski_Metric} so let us study the right diamond \region . There we define the new coordinates
\begin{equation}
 \sigma^\pm \equiv \pm\frac{1}{\lambda}\ln(\pm\lambda x^\pm)
 \,, \quad
 -\infty < \sigma^\pm < +\infty
 \,,
\end{equation}
such that the metric becomes $\d s^2 = -\d\sigma^+\d\sigma^-$, which is simply the flat Minkowski metric in null coordinates. Therefore we have indeed recovered the Minkowski spacetime by taking $M = 0$ in the CGHS solution \eqref{CGHS_OmegaSolution}. Note that in these $\sigma^\pm$ coordinates the Killing vector $\xi$ is given by $\xi = \partial_+ + \partial_-$, so that if we define the usual Cartesian coordinates $(t,x)$ by $\sigma^\pm = t\pm x$ we get $\xi = \partial_t$, recovering the timelike Killing vector field of the Minkowski spacetime.
\subsection{Classical black hole}
Let us now study the solution \eqref{CGHS_OmegaSolution} for a positive mass $M >0$, which corresponds to the well-known classical black hole of the CGHS model.
\paragraph{Horizon:} as $f=0$ here, $\xi$ is a Killing vector and there is a Killing horizon where $\xi^2=0$. Since
\begin{equation}
 \xi^2 = -\frac{1}{\lambda^2}(\nabla\phi)^2 = \frac{\lambda^2x^+x^-}{\frac{M}{\lambda} - \lambda^2x^+x^-}
 \,,
\end{equation}
this horizon corresponds to the two axes $x^\pm=0$. The sign of $\xi^2$ changes when crossing the horizon so it is also an apparent horizon, $\xi$ is timelike outside and spacelike inside. This is the CGHS black hole event horizon.
\paragraph{Singularity:}the scalar curvature is given by
\begin{equation}
 R = \frac{4M\lambda^2}{M-\lambda^3x^+x^-}
 \,,
 \label{CGHS_Static_Curvature}
\end{equation}
so there is a curvature singularity located on the curve $(x^+_s,x^-_s)$ defined by
\begin{equation}
 x^+_sx^-_s = \frac{M}{\lambda^3}
 \,,
\end{equation}
which corresponds to the boundary of the region $e^{-2\phi} > 0$. Note that, as one should expect, the scalar curvature is everywhere zero for $M=0$ which, as we have seen previously, corresponds to the flat Minkowski spacetime. Using the criteria established in Appendix \ref{Appendix_TimelikeSpacelike} with $F(x^+,x^-) = x^+x^- - \frac{M}{\lambda^3}$ we get
\begin{equation}
 \frac{\partial_+F}{\partial_-F} = \frac{x^-}{x^+} = \frac{M}{\lambda^3(x^+)^2} > 0
 \,,
\end{equation}
which means that this singularity is spacelike.

\paragraph{Geodesic (in)completeness:} now let us check whether geodesics are complete when approaching the singularity. To do this (see Appendix \ref{AppendixConformalGeodesics}) we first take a null geodesic defined by $x^+ = x^+_0>0$ and look at the integral
\begin{equation}
 \Delta\chi = \int_{x^-_0}^{x^-}\d\tilde{x}^-\,e^{2\phi(x^+_0,\tilde{x}^-)}
 \,,
\end{equation}
in the limit $x^- \rightarrow \frac{M}{\lambda^3x^+_0}$. We have
\begin{equation}
 \Delta\chi = \int_{x^-_0}^{x^-}\frac{\lambda\,\d\tilde{x}^-}{M - \lambda^3x^+_0\tilde{x}^-}
 = -\frac{1}{\lambda^2x^+_0}\Bigl[\ln(M-\lambda^3x^+_0\tilde{x}^-)\Bigr]_{x^-_0}^{x^-}
 \,,
\end{equation}
which diverges when $x^- \rightarrow \frac{M}{\lambda^3x^+_0}$. Therefore null geodesics are complete when approaching the singularity of the CGHS black hole.

For timelike geodesics we use static coordinates (see appendices \ref{AppendixCoordinateTransition} and \ref{AppendixTimelikeGeodesics}) where for a timelike geodesic going from $\phi_0$ to $\phi$ the change in proper time $\tau$ is
\begin{equation}
 \Delta\tau = \int_{\phi_0}^\phi \d\tilde{\phi}\,\sqrt{\frac{h^2(\tilde{\phi})}{E^2-g(\tilde{\phi})}}
 \,, \; E \in \mathbb{R}
 \,,
\end{equation}
with
\begin{equation}
 g(\phi) = e^{2\phi}Z(\phi) \,, \quad
 h(\phi) = \frac{1}{2\lambda}Z'(\phi) \,, \quad
 Z(\phi) = e^{-2\phi} + \frac{M}{\lambda}
 \,.
\end{equation}
The singularity is located at $\phi \rightarrow +\infty$ and it is easy to check that this integral is convergent in this limit, which means that timelike geodesics are incomplete when approaching the singularity (the geodesic incompleteness in Dilaton gravity models has been studied in the past. See e.g. \cite{Katanaev:1996ni}).

\begin{figure}[htb]
 \centering
 \includegraphics[scale=1]{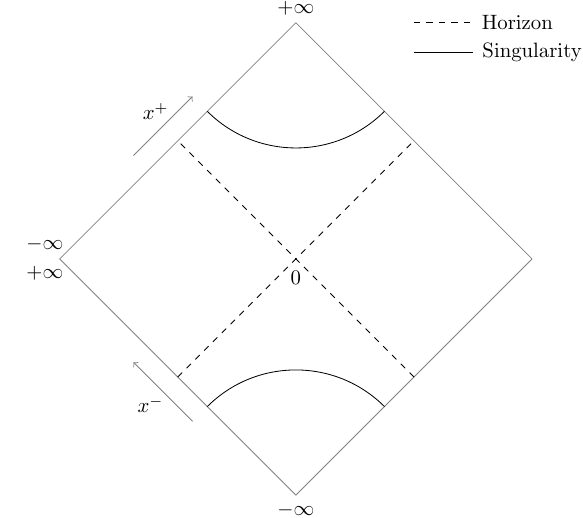}
 \caption{ Static CGHS spacetime for $f=0$ and positive mass $M > 0$. Both singularities are spacelike and hidden behind a Killing apparent horizon. The spacetime is asymptotically flat and not geodesically complete. This picture also applies to the spacetime for the Hartle-Hawking state in the RST model.}
 \label{CGHS_StaticSpacetime}
\end{figure}

\paragraph{Asymptotic flatness:} at spatial and null infinity we have $x^\pm \rightarrow \pm \infty$ and using \eqref{CGHS_Static_Curvature} we get that  the scalar curvature $R\rightarrow 0$ in this limit so the spacetime is asymptotically flat. We actually recover the Minkowski spacetime asymptotically as in this limit we have $e^{-2\phi} \sim -\lambda^2 x^+x^-$ which was the solution for $M=0$. We can define the asymptotic coordinates $\sigma^\pm = \frac{1}{\lambda}\ln(\pm\lambda x^\pm)$ to put the metric under the flat form $\d s^2 = -\d\sigma^+\d\sigma^-$ at infinity. This spacetime is represented on figure \ref{CGHS_StaticSpacetime}.

\subsection{Perturbed classical black hole}
\begin{figure}[htb]
 \centering
 \includegraphics[scale=1]{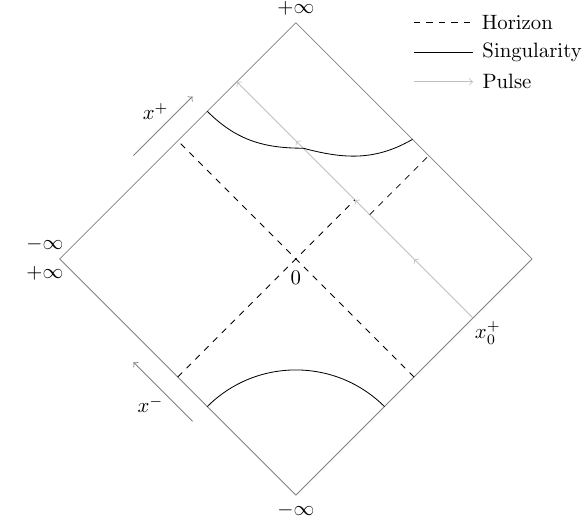}
 \caption{CGHS spacetime perturbed by a shock wave of classical matter sent in at $x^+=x^+_0$. It changes the trajectory of the singularity, which remains spacelike, and the position of the horizon. As for the static solution this picture also applies to the perturbed Hartle-Hawking spacetime of the RST model.}
 \label{CGHS_DynamicSpacetime}
\end{figure}

Now that we have studied the static solution for the CGHS model we would like to know what happens when it is perturbed by infalling classical matter. Equation \eqref{CGHS_u+-Equation} can be integrated twice to get that, up to constant translations of $x^\pm$ and of the matter distribution $\partial_\pm f$,
\begin{equation}
 u_\pm(x^\pm) = C_\pm  - \frac{1}{2}\int_{-\infty}^{x^\pm}\d u\, \int_{-\infty}^{u}\d v\, \big(\partial_\pm f(v)\big)^2
 \,, \quad
 C_\pm \in \mathbb{R}
 \,.
 \label{CGHS_u+-Solution}
\end{equation}
For simplicity let us consider a pulse of mass $m$ traveling along a null geodesic $x^+=x^+_0$ in the $x^-$ direction. The corresponding stress-energy tensor is
\begin{equation}
 \frac{1}{2}(\partial_+f)^2 = \frac{m}{\lambda x^+_0}\delta(x^+-x^+_0)
 \,.
 \label{CGHS_MatterStressEnergyDelta}
\end{equation}
This normalization guarantees that the mass of the spacetime will increase by the mass $m$ of the shock wave. Therefore $u_- = C_-$ is simply a constant while
\begin{equation}
 u_+(x^+) = C_+ - \frac{m}{\lambda x^+_0}\int_{-\infty}^{x^+}\d u\,\int_{-\infty}^u\d v\, \delta(v-x^+_0)
 \,,
\end{equation}
where
\begin{equation}
 \int_{-\infty}^u\d v\, \delta(v-x^+_0)
 = \theta(u - x^+_0)
 \,,
\end{equation}
with $\theta$ the Heaviside function. Therefore
\begin{equation}
 u_+(x^+) = C_+
 - \frac{m}{\lambda x^+_0}(x^+-x^+_0)\theta(x^+-x^+_0)
 \,.
\end{equation}
Hence, defining $\frac{M}{\lambda} \equiv C_+ + C_-$ to recover the previous solution for $x^+ < x^+_0$,  the solution is given by
\begin{equation}
 e^{-2\phi} = \frac{M}{\lambda} - \lambda^2x^+x^- -
 \frac{m}{\lambda x^+_0}(x^+-x^+_0)\theta(x^+-x^+_0)
 \,.
\end{equation}
The spacetime before the shock wave ($x^+ < x^+_0$) is the static CGHS solution, with a singularity hidden behind an horizon. After the shock wave ($x^+ > x^+_0$), the solution becomes
\begin{equation}
 e^{-2\phi} = \frac{M+m}{\lambda} -\lambda^2x^+\biggl(x^-+\frac{m}{\lambda^3x^+_0}\biggr)
 \label{dilatonPhi}
 \,,
\end{equation}
which is exactly the same as \eqref{CGHSConformalStaticSolution} with total mass $M+m$ and a translation $x^- \rightarrow x^- - \frac{m}{\lambda^3x^+_0}$. The event horizon is shifted from $x^-=0$ to $x^- = -\frac{m}{\lambda^3x^+_0}$ and
the singularity lies on the curve $(x^+_s, x^-_s)$ defined by
\begin{equation}
 x^+_s\bigg(x^-_s + \frac{m}{\lambda^3x^+_0}\bigg)
 = \frac{M+m}{\lambda^3}
 \,,
\end{equation}
so the trajectory of the singularity is continuous at $x^+=x^+_0$, but its derivative is not. This dynamical spacetime is represented on figure \ref{CGHS_DynamicSpacetime}.

\section{The RST model}\label{sec:RST}

The RST model \cite{russo_endpoint_1992} is an extension of the CGHS model discussed previously. It takes into account the back-reaction of conformal matter on the geometry of spacetime by including the quantum conformal anomaly into the action. 
Since the respective theory is non-local one has to specify certain boundary conditions. This effectively reduces to a choice of the quantum state. We reserve the possibility to have a situation when different quantum fields are in different quantum states, as was discussed in Introduction. 
We consider the following action
\begin{equation}
 \mathcal{S} = \mathcal{S}_0 + \mathcal{S}_1 + \mathcal{S}_2
 \,,
\end{equation}
where $\mathcal{S}_0$ is the CGHS action \eqref{CGHS_Action} and
\begin{equation}
 \mathcal{S}_1 = -\sum_{i=1}^2\frac{\kappa_i}{2\pi}\int\d ^2x\,\sqrt{-g}\bigg\{\frac{1}{2}(\nabla\psi_i)^2 + \psi_i R\bigg\}
 \,,
 \label{RST_Action1}
\end{equation}
\begin{equation}
 \mathcal{S}_2 = - \frac{\kappa_1+\kappa_2}{2\pi}\int\d ^2x\,\sqrt{-g}\phi R
 \,.
 \label{RST_Action2}
\end{equation}
$\mathcal{S}_1$ is the Polyakov action written under its local form with two auxiliary fields $\psi_1$ and $\psi_2$ so that we can study situations where two different types of particles are present, each having different boundary conditions. The constants $\kappa_1$ and $\kappa_2$ are the central charges associated with $\psi_1$ and $\psi_2$, respectively. The central charge is taken to be positive for physical particles and negative for non-physical ones (such as ghosts). In order to recover the usual RST model one simply has to take $\kappa_2=0$. First we will compute a general solution, before studying various particular cases.
\subsection{Equations of motion}
For each action term $\mathcal{S}_i$ we define the associated stress-energy tensor by
\begin{equation}
 T^{(i)}_{\mu\nu} \equiv -\frac{2}{\sqrt{|g|}}\frac{\delta \mathcal{S}_i}{\delta g^{\mu\nu}}
 \,,
\end{equation}
and as this is a two-dimensional gravity model the total stress-energy tensor must vanish, \textit{i.e.}
\begin{equation}
 T_{\mu\nu} \equiv T_{\mu\nu}^{(0)} + T_{\mu\nu}^{(1)} + T_{\mu\nu}^{(2)} = 0
 \,.
\end{equation}
Of course $T^{(0)}_{\mu\nu}$ is the same as in the CGHS model
\begin{equation}
 T^{(0)}_{\mu\nu}
 = \frac{2}{\pi}e^{-2\phi}\Big\{-\nabla_\mu\nabla_\nu\phi + g_{\mu\nu}\big[\square\phi - (\nabla\phi)^2 + \lambda^2\big]\Big\}
 - \frac{1}{4\pi}g_{\mu\nu}(\nabla f)^2
 + \frac{1}{2\pi}\partial_\mu f \partial_\nu f
 \,,
 \label{RST_StressEnergy0}
\end{equation}
while for $\mathcal{S}_1$ and $\mathcal{S}_2$ we get
\begin{equation}
 T^{(1)}_{\mu\nu} = \frac{1}{\pi}\sum_{\kappa=1}^2\kappa_i\bigg\{\frac{1}{2}\partial_\mu\psi_i\partial_\nu\psi_i
 -\nabla_\mu\nabla_\nu\psi_i
 +g_{\mu\nu}\bigg(\square\psi_i - \frac{1}{4}(\nabla\psi_i)^2\bigg)\bigg\}
 \,,
 \label{RST_StressEnergy1}
\end{equation}
and
\begin{equation}
 T_{\mu\nu}^{(2)} = \frac{1}{\pi}(\kappa_1+\kappa_2)(g_{\mu\nu}\square\phi - \nabla_\mu\nabla_\nu\phi)
 \,.
 \label{RST_StressEnergy2}
\end{equation}
Each auxiliary field $\psi_i$ satisfies
\begin{equation}
 \square\psi_i = R
 \,, \quad
 i=1,2
 \,,
 \label{RST_PsiEquation}
\end{equation}
so that by plugging this into \eqref{RST_Action1} one gets the usual Polyakov action. The equation for the dilaton becomes
\begin{equation}
 R\bigg(1 + \frac{\kappa_1+\kappa_2}{2}e^{2\phi}\bigg) + 4\big\{\square\phi - (\nabla\phi)^2 + \lambda^2\big\} = 0
 \,,
 \label{RST_DilatonEquation}
\end{equation}
and the equation for the matter is unchanged
\begin{equation}
 \square f = 0
 \,.
 \label{RST_MatterEquation}
\end{equation}
Finally the trace of the stress-energy tensor is
\begin{equation}
 g^{\mu\nu}T_{\mu\nu} =
 \frac{2}{\pi}e^{-2\phi}\Big\{\square\phi - 2(\nabla\phi)^2 + 2\lambda^2\Big\} + \frac{1}{\pi}\sum_{i=1}^2\kappa_i\square(\psi_i+\phi) = 0
 \,,
\end{equation}
and using \eqref{RST_PsiEquation} to combine this with \eqref{RST_DilatonEquation} so as to eliminate the terms containing $(\nabla\phi)^2$ we get that
\begin{equation}
 (R+2\square\phi)(\kappa_1 + \kappa_2 - 2e^{-2\phi}) = 0
 \,.
\end{equation}
This implies that, for a non-constant dilaton,
\begin{equation}
 R + 2\square\phi = 0
 \,,
 \label{RST_Trace}
\end{equation}
which is the same condition as in the CGHS model \eqref{CGHS_Trace}.

The other possible solution to (\ref{RST_Trace}) is to have a constant dilaton field $\phi=-\frac{1}{2}\ln (\frac{\kappa_1+\kappa_2}{2})$.
It corresponds to a 2d spacetime with constant curvature $R=-2\lambda^2$. We do not consider this solution here.

\subsection{Conformal gauge}
Just like in the CGHS model, the condition \eqref{RST_Trace} expressed in the conformal gauge imposes that
\begin{equation}
 \partial_+\partial_-(\rho-\phi) = 0
 \,,
\end{equation}
so
\begin{equation}
 2\rho = 2\phi + j
 \,, \quad
 j = j_+(x^+) + j_-(x^-)
 \,,
\end{equation}
and once again we can perform a change of coordinates defined by
\begin{equation}
 \tilde{x}^\pm = \int e^{j_\pm(x^\pm)}\d x^\pm
 \,,
\end{equation}
to set $\rho=\phi$.
In these new coordinates the components of the stress-energy tensor are given by \eqref{CGHS_ConformalStressEnergy} and
\begin{equation}
 \left\{
 \begin{aligned}
  T^{(1)}_{\pm\pm} & = \frac{1}{\pi}\sum_{i=1}^2\kappa_i\bigg\{\frac{1}{2}(\partial_\pm\psi_i)^2 - \partial_\pm^2\psi_i + 2\partial_\pm\phi \, \partial_\pm\psi_i\bigg\}
  \,, \\
  T^{(1)}_{+-} & = \frac{1}{\pi}\sum_{i=1}^2\kappa_i\partial_+\partial_-\psi_i
  \,.
 \end{aligned}
 \right.
\end{equation}
and
\begin{equation}
 \left\{
 \begin{aligned}
  T^{(2)}_{\pm\pm} & = \frac{1}{\pi}(\kappa_1+\kappa_2)\big\{2(\partial_\pm\phi)^2 - \partial_\pm^2\phi\big\}
  \,, \\
  T^{(2)}_{+-} & = \frac{1}{\pi}(\kappa_1+\kappa_2)\partial_+\partial_-\phi
  \,.
 \end{aligned}
 \right.
\end{equation}
Combining equations \eqref{RST_PsiEquation} and \eqref{RST_Trace} we obtain that $\square(\psi + 2\phi) = 0$ so each auxiliary field $\psi_i$ can be expressed as
\begin{equation}
 \psi_i = -2\phi + w_i
 \,,
\end{equation}
with $\square w_i = 0$, \textit{i.e.}
\begin{equation}
 w_i = w_{i+}(x^+) + w_{i-}(x^-)
 \,.
\end{equation}
The functions $w_i$ will be determined by the boundary conditions of each auxiliary field $\psi_i$. Using this the components of $T^{(1)}_{\mu\nu}$ become
\begin{equation}
 \left\{
 \begin{aligned}
  T^{(1)}_{\pm\pm} & = \frac{2}{\pi}\sum_{i=1}^2\kappa_i\bigl\{\partial_\pm^2\phi - (\partial_\pm\phi)^2 - t_{i\pm}\bigr\}
  \,, \\
  T^{(1)}_{+-} & = -\frac{2}{\pi}(\kappa_1 + \kappa_2)\partial_+\partial_-\phi
  \,,
 \end{aligned}
 \right.
\end{equation}
where we have defined
\begin{equation}
 t_{i\pm} \equiv \frac{1}{2}w_{i\pm}'' - \frac{1}{4}(w_{i\pm}')^2
 \,.
 \label{EQ_Definition_t12}
\end{equation}
We also define $T^{(12)}_{\mu\nu} \equiv T^{(1)}_{\mu\nu} + T^{(2)}_{\mu\nu}$, which we consider as the stress-energy tensor of the quantum matter, while $T^{(0)}_{\mu\nu}$ is the purely geometrical part. It is given by
\begin{equation}
 \left\{
 \begin{aligned}
  T^{(12)}_{\pm\pm} & = \frac{1}{\pi}\sum_{i=1}^2\kappa_i\bigl\{\partial_\pm^2\phi - 2t_{i\pm}\bigr\}
  \,, \\
  T^{(12)}_{+-} & = -\frac{1}{\pi}(\kappa_1 + \kappa_2)\partial_+\partial_-\phi
  \,.
 \end{aligned}
 \right.
 \label{RST_T12Definition}
\end{equation}
The $\pm\pm$ component of the total stress energy tensor is now
\begin{equation}
 T_{\pm\pm} = \frac{1}{\pi}e^{-2\phi}\big\{4(\partial_\pm\phi)^2 -2\partial_\pm^2\phi \big\}
 + \frac{1}{\pi}\sum_{i=1}^2\kappa_i(\partial_\pm^2\phi - 2t_{i\pm})
 + \frac{1}{2\pi}(\partial_\pm f)^2
 = 0
 \,,
\end{equation}
and the $+-$ component gives
\begin{equation}
 T_{+-} = \frac{1}{\pi}(2e^{-2\phi} - \kappa_1-\kappa_2)\partial_+\partial_-\phi - \frac{4}{\pi}e^{-2\phi}\partial_+\phi\partial_-\phi - \frac{\lambda^2}{\pi} = 0
 \,.
\end{equation}
We define the new variable
\begin{equation}
 \Omega(\phi) \equiv (\kappa_1+\kappa_2)\phi + e^{-2\phi}
 \,,
 \label{RST_OmegaDefinition}
\end{equation}
and we get that
\begin{equation}
 \pi T_{\pm\pm} = \partial_\pm^2\Omega -2\sum_{i=1}^2\kappa_i t_{i\pm} + \frac{1}{2}(\partial_\pm f)^2 = 0
 \,,
\end{equation}
and
\begin{equation}
 \pi T_{+-} = -\partial_+\partial_-\Omega - \lambda^2 = 0
 \,.
\end{equation}
Hence $\Omega$ can be expressed as
\begin{equation}
 \Omega = -\lambda^2x^+x^- + u_+(x^+) + u_-(x^-)
 \,,
\end{equation}
Inserting this into the equation for $T_{\pm\pm}$ gives
\begin{equation}
 u_\pm'' = 2\sum_{i=1}^2\kappa_i t_{i\pm} - \frac{1}{2}(\partial_\pm f)^2
 \,.
 \label{RST_u+-_DiffEquation}
\end{equation}
Note that, defining $\Omega' \equiv \frac{\d \Omega}{\d \phi}$, the curvature can be expressed in terms of $\Omega$ as
\begin{equation}
 R = \frac{8e^{-2\phi}}{\Omega'}\bigg(\partial_+\partial_-\Omega - \frac{\Omega''}{\Omega'^2}\partial_+\Omega\partial_-\Omega\bigg)
 \,,
 \label{RST_CurvatureOmega}
\end{equation}
so we will have to look for possible curvature singularities where $\Omega'$ vanishes. We also have
\begin{equation}
 (\nabla\phi)^2 = -4e^{-2\phi}\partial_+\phi\partial_-\phi
 \,,
\end{equation}
and $\partial_\pm\phi = \frac{1}{\Omega'}\partial_\pm\Omega$ so the condition $(\nabla\phi)^2=0$ is satisfied if $\partial_\pm\Omega = 0$. This means that we will have to look for possible apparent horizons or wormhole throats where $\partial_\pm \Omega = 0$.
\subsection{Static solution}
\label{Section_StaticSolutionRST}
For a vector field $\xi = \xi^+ \partial_+ + \xi^-\partial_-$ the Killing equations give
\begin{equation}
 \left\{
 \begin{aligned}
  & \partial_+ \xi^- = 0 = \partial_-\xi^+
  \,, \\
  & \partial_+(\xi^+e^{2\phi}) + \partial_-(\xi^-e^{2\phi}) = 0
  \,.
 \end{aligned}
 \right.
\end{equation}
Assume that $\phi (x^+,x^-) = \tilde{\phi} (x^+x^-)$. We can show that this situation corresponds to a static solution. Indeed in this case the second Killing equation becomes
\begin{equation}
 \partial_+\xi^+ + \partial_-\xi^- + (x^-\xi^+ + x^+\xi^-)\tilde{\phi}' = 0
 \,,
\end{equation}
so the vector $\xi = \lambda(x^+\partial_+ - x^-\partial_-)$ is a Killing vector. Note that this is the same expression as we had found for the Killing vector of the CGHS spacetime \eqref{CGHS_Killing_Coordinates}. As we will see soon, this normalization guarantees that $\xi = \partial_t$ in asymptotically flat coordinates at infinity. Its norm is
\begin{equation}
 \xi^2 = g_{\mu\nu}\xi^\mu\xi^\nu = \lambda^2 e^{2\rho}x^+x^-
 \,,
\end{equation}
so $\xi$ is timelike where $x^+x^- < 0$ and spacelike where $x^+x^- > 0$ and there is a Killing horizon where $x^+x^- = 0$.

For such a static solution, $\Omega$ is also a function of the product $x^+x^-$ and we must have
\begin{equation}
 u_+(x^+) + u^-(x^-) = \tilde{u}(x^+x^-)
 \,,
\end{equation}
which means that, for a solution without matter ($f=0$),
\begin{equation}
 u_\pm(x^\pm) = 2(\kappa_1 P_1 + \kappa_2P_2) \ln\big(\lambda |x^\pm|\big) + C_\pm
 \,, \quad
 P_{1,2},\, C_\pm \in \mathbb{R}
 \,,
\end{equation}
Therefore
\begin{equation}
 \Omega = -\lambda^2x^+x^-
 + 2(\kappa_1 P_1 + \kappa_2P_2)\ln|\lambda^2x^+x^-|
 + \frac{M}{\lambda}
 \,, \quad
 M \geq 0
 \,.
 \label{RST_OmegaExtendedStaticSolution}
\end{equation}
Taking $\kappa_1 = \kappa_2 = 0$ we must recover the CGHS solution \eqref{CGHS_OmegaSolution} where the parameter $M$ corresponds to the ADM mass of the spacetime so we take it non negative. Since $\Omega$ diverges when $x^+x^- \rightarrow 0$ we can expect a coordinate singularity there so let us study this solution on one of the four quadrants delimited by the two axes $x^\pm = 0$. We want the Killing vector $\xi$ to be timelike at infinity so we choose the region \region (which by symmetry is equivalent to the region $\{x^+<0, x^->0\}$). Then
\begin{equation}
 \Omega = -\lambda^2x^+x^-
 + 2(\kappa_1 P_1 + \kappa_2P_2)\ln(-\lambda^2x^+x^-)
 + \frac{M}{\lambda}
 \,.
 \label{RST_OmegaStaticSolution}
\end{equation}
We can recover the solution of the CGHS model \eqref{CGHS_OmegaSolution} by taking $\kappa_1 = \kappa_2 = 0$, while if we take $(P_1,P_2,M) = (-\frac{1}{4},-\frac{1}{4},0)$ then
\begin{equation}
 \kappa\phi + e^{-2\phi} = -\lambda^2x^+x^- - \frac{\kappa}{2}\ln(-\lambda^2x^+x^-)
 \,,
\end{equation}
and the solution for this equation is $e^{-2\phi} = -\lambda^2x^+x^-$ which is the flat Minkowski spacetime, as discussed in section \ref{Section_Minkowski_CGHS}. Taking the limit $x^+x^- \rightarrow -\infty$ with $\phi\rightarrow -\infty$ in \eqref{RST_OmegaStaticSolution} we get that $e^{-2\phi} \sim -\lambda^2x^+x^-$, which means that asymptotically this static solution is also the Minkowski spacetime. Using the asymptotic coordinates $\sigma^\pm = \pm\frac{1}{\lambda}\ln(\pm x^\pm)$ the metric takes the flat form $\d s^2 = -\d\sigma^+\d\sigma^-$ and the Killing vector $\xi$ becomes $\xi = \partial_+ + \partial_- = \partial_t$. We can check explicitly that the curvature goes to zero in this limit by using the formula \eqref{RST_CurvatureOmega} with
\begin{equation}
 \Omega \sim e^{-2\phi} \sim -\lambda^2x^+x^-
 \,, \quad
 \Omega' \sim -2e^{-2\phi}
 \,, \quad
 \Omega'' =4e^{-2\phi}
 \,,
\end{equation}
and
\begin{equation}
 \partial_\pm\Omega = -\lambda^2x^\mp + 2\frac{\kappa_1P_1+\kappa_2P_2}{x^\pm} \sim -\lambda^2x^\mp
 \,, \quad
 \partial_+\partial_-\Omega = -\lambda^2
 \,,
\end{equation}
to get
\begin{equation}
 R \rightarrow -4(-\lambda^2 + \lambda^2) = 0
 \,.
\end{equation}

Finally, inserting \eqref{RST_OmegaStaticSolution} into the $T_{\pm\pm}$ equations gives
\begin{equation}
 u_\pm'' = -\frac{2(\kappa_1 P_1 + \kappa_2P_2)}{(x^\pm)^2} = 2(\kappa_1 t_{1\pm} + \kappa_2 t_{2\pm})
 \,,
\end{equation}
therefore we can take for a general pair $(\kappa_1, \kappa_2)$
\begin{equation}
 t_{i\pm} = - \frac{P_i}{(x^\pm)^2}
 \,.
 \label{RST_t+_StaticSolution}
\end{equation}
The value of the constants $P_i$ depends on the quantum state in which the particles described by the auxiliary fields $\psi_i$ are, as we will discuss soon.
\subsection{Perturbed solution}
Just as in the CGHS model, let us now consider the static solution for the RST model seen previously perturbed by a pulse of classical matter of mass $m$ traveling along a null geodesic in the $x^-$ direction. The corresponding stress-energy tensor is once again
\begin{equation}
 \frac{1}{2}(\partial_+f)^2 = \frac{m}{\lambda x^+_0}\delta(x^+-x^+_0)
 \,,
 \label{RST_MatterStressEnergyDelta}
\end{equation}
and we need to integrate \eqref{RST_u+-_DiffEquation}, using \eqref{RST_t+_StaticSolution}. Imposing that for $x^+<x^+_0$ the solution should be identical to the static situation described by \eqref{RST_OmegaStaticSolution} we get
\begin{equation}
 \Omega = -\lambda^2x^+x^-
 + 2(\kappa_1 P_1 + \kappa_2P_2)\ln(-\lambda^2x^+x^-)
 + \frac{M}{\lambda} - \frac{m}{\lambda x^+_0}(x^+-x^+_0)\theta(x^+-x^+_0)
 \,.
\end{equation}
After the shock, $x^+ > x^+_0$, we have
\begin{equation}
 \Omega = -\lambda^2x^+\biggl(x^-+\frac{m}{\lambda^3x^+_0}\biggr)
 + 2(\kappa_1 P_1 + \kappa_2P_2)\ln(-\lambda^2x^+x^-)
 + \frac{M+m}{\lambda}
 \,,
 \label{RST_OmegaDynamicalSolution}
\end{equation}
which is no longer static as $\Omega$ cannot be written as a function of the product $x^+x^-$. We will later study the effect of this shock wave on the spacetime, according to the quantum states of particles.
\subsection{Quantum states}
\label{SubsectionQuantumStates}
Now we would like to determine which quantum states correspond to which values of the constants $P_i$ in \eqref{RST_t+_StaticSolution}. To do this we consider the usual RST model, that is to say with only one auxiliary field $\psi = \psi_1$ of central charge $\kappa = \kappa_1$ and we take $\kappa_2 = 0$. Then \eqref{RST_OmegaStaticSolution} becomes
\begin{equation}
 \Omega = \kappa\phi + e^{-2\phi} = -\lambda^2x^+x^- + 2\kappa P\ln(-\lambda^2x^+x^-) + \frac{M}{\lambda}
 \,.
 \label{RST_OmegaStaticSolutionKappa2=0}
\end{equation}
We need to compute the energy density detected by an observer at infinity, the quantum part of the stress-energy tensor being
\begin{equation}
 \left\{
 \begin{aligned}
  T^{(12)}_{\pm\pm} & = \frac{\kappa}{\pi}\biggl\{\partial_\pm^2\phi + 2\frac{P}{(x^\pm)^2}\biggr\}
  \,, \\
  T^{(12)}_{+-} & = -\frac{\kappa}{\pi}\partial_+\partial_-\phi
  \,.
 \end{aligned}
 \right.
\end{equation}
Therefore let us study these quantities when $x^+x^- \rightarrow -\infty$ with $\phi \rightarrow -\infty$. In this limit $e^{-2\phi} \sim -\lambda^2x^+x^-$ and a careful analysis shows that
\begin{equation}
 \partial_\pm^2\phi \sim \frac{1}{2(x^\pm)^2}
 \,, \quad
 \partial_+\partial_-\phi \sim -\frac{\kappa P}{\lambda^2(x^+x^-)^2}
 \,,
\end{equation}
which leads directly to
\begin{equation}
 \left\{
 \begin{aligned}
  T^{(12)}_{\pm\pm} & \sim \frac{\kappa}{\pi}\frac{1+4P}{2(x^\pm)^2}
  \,, \\
  T^{(12)}_{+-} & \sim \frac{\kappa^2 P}{\pi\lambda^2(x^+x^-)^2}
  \,.
 \end{aligned}
 \right.
\end{equation}
Then we need to compute these components in the coordinates $\sigma^\pm = \pm\frac{1}{\lambda}\ln(\pm\lambda x^\pm)$ in which the metric is asymptotically flat, and we get
\begin{equation}
 \left\{
 \begin{aligned}
  \tilde{T}^{(12)}_{\pm\pm} & = \biggl(\frac{\d x^\pm}{\d\sigma^\pm}\biggr)^2 T^{(12)}_{\pm\pm} \sim \frac{\lambda^2 \kappa}{2\pi}(1+4P)
  \,, \\
  \tilde{T}^{(12)}_{+-} & = \frac{\d x^+}{\d\sigma^+}\frac{\d x-}{\d\sigma^-}T^{(12)}_{+-} \sim -\frac{\kappa^2 P}{\pi x^+x^-} \rightarrow 0
  \,.
 \end{aligned}
 \right.
\end{equation}
Thus the radiation detected by an observer at infinity will be zero only for $P=-\frac{1}{4}$, which therefore corresponds to the Boulware state. Note that the Minkowski spacetime, for which $P=-\frac{1}{4}$ and $M=0$ as discussed previously, is simply the Boulware state with zero mass. This was to be expected as there should obviously be no radiation in Minkowski spacetime.

Then for $P=0$ we get $T^{(12,\sigma)}_{\pm\pm} \rightarrow \frac{\lambda^2 \kappa}{2\pi}$ which is the energy density of a thermal gas with temperature $T = \frac{\lambda}{2\pi}$. Besides for $P=0$ the solution \eqref{RST_OmegaStaticSolutionKappa2=0} is regular at $x^+x^-=0$ and can be extended throughout the whole plane $\{-\infty < x^\pm < +\infty\}$. Therefore there is a Killing horizon  where $x^+x^-=0$ and $P=0$ corresponds to the Hartle-Hawking state.

Note that these values of the constant $P$ for the Hartle-Hawking and Boulware quantum states are consistent with what was found in \cite{potaux_quantum_2022} (see Appendix \ref{AppendixCoordinateTransition} for a detailed discussion).

\section{Single quantum states}\label{sec:singlestate}

In this section we will consider the usual RST model, that is to say when all the particles, described by a single auxiliary field $\psi$ and with central charge $\kappa$, are in the same quantum state. Therefore we use the solution \eqref{RST_OmegaStaticSolutionKappa2=0}, taking $P=0$ for the Hartle-Hawking state and $P=-\frac{1}{4}$ for the Boulware state. As noted in section \ref{Section_StaticSolutionRST} the static solutions thus obtained will all be asymptotically flat. The dynamical solutions will be obtained by taking \eqref{RST_OmegaDynamicalSolution} with $(\kappa_1,P_1) = (\kappa,P)$ and $\kappa_2=0$.

\subsection{Hartle-Hawking state}
We start by examining the case where the particles are in the Hartle-Hawking state ($P=0$) \cite{Solodukhin:1995te}, meaning in particular that thermal radiation can be detected at infinity. Since non-physical particles should not be observed at infinity we will only consider physical particles here, so we take $\kappa > 0$.
\subsubsection{Static solution}
In this situation the static solution \eqref{RST_OmegaStaticSolutionKappa2=0} becomes
\begin{equation}
 \Omega = \kappa\phi + e^{-2\phi} = -\lambda^2x^+x^- + \frac{M}{\lambda}
 \,, \quad
 M \geq 0
 \,,
 \label{RST_OmegaStaticSolutionHH}
\end{equation}
and it is valid on the whole plane $\{-\infty < x^\pm < +\infty\}$. Let us study the structure of this spacetime.
\paragraph{Horizon:} as discussed while defining the Hartle-Hawking state in section \ref{SubsectionQuantumStates} this spacetime has a Killing horizon where $x^\pm=0$ which coincides with the apparent horizon condition $\partial_\pm\Omega = 0$ since $\partial_\pm\Omega = -\lambda^2x^\mp$ here.
\paragraph{Singularity:} since $\kappa > 0$ here $\Omega' = \kappa - 2e^{-2\phi}$ vanishes at $\phi = \phi_{cr} \equiv -\frac{1}{2}\ln\frac{\kappa}{2}$ where $\Omega$ takes the value
\begin{equation}
 \Omega_{cr} \equiv \Omega(\phi_{cr}) = \frac{\kappa}{2}\biggl(1-\ln\frac{\kappa}{2}\biggr)
 \,.
 \label{RST_OmegaCritical}
\end{equation}
Therefore, according to \eqref{RST_CurvatureOmega} there is a curvature singularity on the curve $(x^+_s, x^-_s)$ defined by $\Omega(x^+_s, x^-_s) = \Omega_{cr}$ \textit{i.e.}
\begin{equation}
 x^+_s x^-_s
 = \frac{M}{\lambda^3} - \frac{\kappa}{2\lambda^2}\bigg(1-\ln\frac{\kappa}{2}\bigg)
 \,.
 \label{RST_StaticSingularityHH}
\end{equation}
This singularity is located behind the horizon $x^+x^-=0$ if and only if the right-hand-side of this equation is positive, \textit{i.e.}
\begin{equation}
 M > M_{cr} \equiv \frac{\lambda\kappa}{2}\bigg(1-\ln\frac{\kappa}{2}\bigg)
 \,.
 \label{RST_CriticalMassHH}
\end{equation}
If this is the case then the singularity is spacelike, as in the classical black hole.
As for the CGHS solution, we can study whether null geodesics are complete near the singularity by considering the integral
\begin{equation}
 \Delta\chi = \int_{x^-_0}^{x^-_s(x^+_0)}\d\tilde{x}^-\,e^{2\phi(x^+_0\tilde{x}^-)}
 \,,
\end{equation}
where $x^-_s(x^+_0)$ defined by \eqref{RST_StaticSingularityHH} is the point where the geodesic meets the singularity. Since at this point $e^{-2\phi}$ takes the finite value $e^{-2\phi_{cr}} = \frac{\kappa}{2}$ this integral is convergent so the geodesic is incomplete. Note that this is a difference with the classical CGHS singularity for which null geodesics were complete.  The finite value of $\phi$ on the singularity also means that the change in proper time for timelike geodesics is finite, so that they are also incomplete.

\bigskip

\noindent{\bf Two-branch space-time.}  The above mentioned incompleteness of null geodesics may have the following interesting  interpretation.  As it was suggested in \cite{Solodukhin:1995te}
the dilaton $\phi$ is a two-fold function of $x^+x^-$. The one branch is valid for values $\phi> \phi_{cr}$ and asymptotically approaches the classical CGHS black hole solution.
The other branch is for $\phi<\phi_{cr}$. The second branch is also asymptotically flat and it contains a Killing horizon. The two branches are glued together at $\phi=\phi_{cr}$, where the
metric is of the ${\cal C}^1$-class. (Notice that the classical CGHS  metric is not even of ${\cal C}^0$-class at the singularity.) This means that the geodesics may not end at the
singularity but go further and re-emerge in the second branch. So that the full two-branch spacetime would be geodesically complete. 
The RST model can be viewed as a one-loop effective action. It was speculated in \cite{Solodukhin:1995te} that taking into account the higher loops 
would smoothen the classical singularity even more so that the two-branch spacetime would be regular but keeping the discussed two-fold structure (see also \cite{Mikovic:1996bh} for some related work on two-loop effects extending the RST model).

\bigskip

Considering the spacetime that corresponds to the single branch $\phi>\phi_{cr}$  we see that the picture of the Hartle-Hawking spacetime for $M > M_{cr}$ is thus pretty much identical to the CGHS solution represented on figure \ref{CGHS_StaticSpacetime}. For $M = M_{cr}$ the singularity coincides with the horizon at $x^+x^-=0$ while for $M < M_{cr}$ there is a naked singularity.
\subsubsection{Dynamical solution}
The dynamical solution for the Hartle-Hawking state ($P=0$) is given by
\begin{equation}
 \Omega = \kappa\phi + e^{-2\phi} = -\lambda^2x^+x^- + \frac{M}{\lambda} - \frac{m}{\lambda x^+_0}(x^+-x^+_0)\theta(x^+-x^+_0)
 \,, \;
 M \geq 0
 \,,
\end{equation}
where we also only consider the case $\kappa > 0$. After the shock wave $(x^+>x^+_0$) the solution can be written as
\begin{equation}
 \Omega = -\lambda^2x^+\tilde{x}^- + \frac{M+m}{\lambda}
 \,, \quad
 \tilde{x}^- \equiv x^- + \frac{m}{\lambda^3x^+_0}
 \,.
\end{equation}
This is the same as \eqref{RST_OmegaStaticSolutionHH} with an increase of mass by $m$ and a translation by $-\frac{m}{\lambda^3x^+_0}$ in the $x^-$ direction. Therefore the structure is essentially the same, the only differences being that the $x^-=0$ horizon is displaced to $x^-_h = -\frac{m}{\lambda^3x^+_0}$ and the singularity lies on the curve $(x^+_s,x^-_s)$ defined by
\begin{equation}
 x^+_s \biggl(x^-_s + \frac{m}{\lambda^3x^+_0}\biggr)
 = \frac{M+m}{\lambda^3} - \frac{\kappa}{2\lambda^2}\bigg(1-\ln\frac{\kappa}{2}\bigg)
 \,.
\end{equation}
If the initial mass $M$ is bigger than the critical mass $M_{cr}$ defined in \eqref{RST_CriticalMassHH} then the singularity will remain hidden behind an horizon and spacelike. As for the static case the picture of the spacetime for $M > M_{cr}$ is identical to the dynamical CGHS solution represented on figure \ref{CGHS_DynamicSpacetime}. 
\subsection{Boulware state: non-physical particles}
We will now study the Boulware state ($P=-\frac{1}{4}$) for which there is no radiation at spatial infinity. A limited number of aspects for Boulware in the context of the RST model was studied in the past in \cite{Zaslavskii:2006pn}. This state is particularly adapted to the non-physical particles ($\kappa < 0$), which should not be detected at infinity, so we will study this situation here.
\subsubsection{Static solution}
Taking $P=-\frac{1}{4}$ in \eqref{RST_OmegaStaticSolution} we get
\begin{equation}
 \Omega = \kappa\phi + e^{-2\phi} = -\lambda^2x^+x^- - \frac{\kappa}{2}\ln(-\lambda^2x^+x^-) + \frac{M}{\lambda}
 \,, \;
 \kappa < 0
 \,,
 \label{RST_OmegaStaticSolutionB_NP}
\end{equation}
and we analyze the structure of this solution on the region \region, following similar steps to what we have done for the Hartle-Hawking state previously.
\paragraph{Singularity:} since $\kappa < 0$ we have $\Omega' = \kappa - 2e^{-2\phi} < 0$ so it never vanishes. Therefore, according to \eqref{RST_CurvatureOmega}, the only possible curvature singularity is on the border of the region of interest, \textit{i.e.}\ where $x^+x^-=0$. Let us therefore study the curvature \eqref{RST_CurvatureOmega} in the limit $x^+x^- \rightarrow 0$. In this limit $\Omega \rightarrow -\infty$ which implies that $\phi \rightarrow +\infty$. Thus
\begin{equation}
 \kappa\phi + \frac{\kappa}{2}\ln(-\lambda^2x^+x^-) - \frac{M}{\lambda} = -e^{-2\phi} - \lambda^2x^+x^- \rightarrow 0
 \,,
\end{equation}
so
\begin{equation}
 e^{\kappa\phi} \sim e^{M/\lambda}(-\lambda^2x^+x^-)^{\kappa/2}
 \,,
\end{equation}
and
\begin{equation}
 e^{-2\phi} \sim e^{-2M/\lambda\kappa}(-\lambda^2x^+x^-) \rightarrow 0
 \,.
 \label{RST_BNP_E-2phiequivalent}
\end{equation}
Note that the behavior or $e^{-2\phi}$ in this limit will be useful when discussing geodesic completeness. Besides
\begin{equation}
 \partial_+\Omega\partial_-\Omega = \frac{1}{x^+x^-}\biggl(-\lambda^2x^+x^- - \frac{\kappa}{2}\biggr)^2 \sim \frac{\kappa^2}{4x^+x^-}
 \,, \quad
 \partial_+\partial_-\Omega = -\lambda^2
 \,,
\end{equation}
and
\begin{equation}
 \Omega' = \kappa - 2e^{-2\phi} \rightarrow \kappa
 \,, \quad
 \Omega'' = 4e^{-2\phi}
 \,.
\end{equation}
Plugging all this into \eqref{RST_CurvatureOmega} we get that $R \rightarrow 0$ in this limit. Hence there is no curvature singularity on the border $x^+x^-=0$ but spacetime is asymptotically flat there.
\paragraph{Geodesic completeness:}now that we know that the region \region is singularity free let us study whether it is geodesically complete. For null geodesics (see Appendix \ref{AppendixConformalGeodesics}) we have to look at the integral
\begin{equation}
 \Delta\chi = \int_{x^-_0}^0\mathrm{d}x^-\,e^{2\phi(x^+_0,x^-)}
 \,,
\end{equation}
but since $e^{2\phi(x^+_0,x^-)} \sim \frac{e^{2M/\lambda\kappa}}{-\lambda^2x^+_0x^-}$ when $x^- \rightarrow 0$ this integral is divergent and null geodesics are complete. Conducing the analysis in static coordinates one can show that timelike geodesics are also complete. Therefore this spacetime is asymptotically flat and geodesically complete at both ends $\phi \rightarrow \pm \infty$.
\paragraph{Horizon:}the apparent horizon condition $\partial_\pm\Omega(x^+_h,x^-_h)=0$ gives the equation
\begin{equation}
 x^+_hx^-_h = -\frac{\kappa}{2\lambda^2} > 0
 \,,
\end{equation}
but this is outside of the region of interest, which thus does not contain any horizon.

\paragraph{Wormhole structure:}Since this spacetime is singularity free and asymptotically flat we can study whether it has the structure of a type II wormhole, as defined at the beginning of section \ref{Section_ClassicalModel}. In order to do this we study the metric function $g(x^+x^-) = -\xi^2 = \lambda^2e^{2\phi}x^+x^-$ to see if it has a minimum for some value of $x^+x^-$. Since $\kappa < 0$ the function $\Omega(\phi)$ is monotonous and so is $\Omega(x^+x^-)$, meaning that for each value of $x^+x^-$ there is a single corresponding value of $\phi$. Then one can check that $g \rightarrow 1$ from below when $x^+x^- \rightarrow -\infty$ and using \eqref{RST_BNP_E-2phiequivalent} we get $g \rightarrow e^{2M/\lambda\kappa} < 1$ (as $\kappa<0$) when $x^+x^- \rightarrow 0$. Therefore the function $g$ does not necessarily have a minimum, and if we impose the conditions $\partial_\pm g= 0$ we get $x^\pm\partial_\pm \phi= -\frac{1}{2}$, which, inserted into the differentiated master equation \eqref{RST_OmegaStaticSolutionB_NP} gives $e^{-2\phi} = -\lambda^2x^+x^-$. But this is only true in the limit $x^+x^- \rightarrow -\infty$, so the function $g$ has no minimum and the spacetime does not have the structure of a black hole mimicker.
\subsubsection{Dynamical solution}
The corresponding dynamical solution is given by
\begin{equation}
 \Omega = \kappa \phi + e^{-2\phi}
 = -\lambda^2x^+x^- - \frac{\kappa}{2}\ln(-\lambda^2x^+x^-)
 + \frac{M}{\lambda} - \frac{m}{\lambda x^+_0}(x^+-x^+_0)\theta(x^+-x^+_0)
 \,, \;
 \kappa < 0
 \,,
\end{equation}
which for $x^+ > x^+_0$ becomes
\begin{equation}
 \Omega = -\lambda^2x^+\tilde{x}^- - \frac{\kappa}{2}\ln(-\lambda^2x^+x^-) + \frac{M+m}{\lambda}
 \,, \;
 \tilde{x}^- = x^- + \frac{m}{\lambda^3x^+_0}
 \,.
\end{equation}
The limit $x^\pm \rightarrow 0$ is essentially the same as in the static case, so the region $\{x^+>0, x^-<0\}$ is still singularity free. The only difference is the appearance of an apparent horizon as the condition $\partial_+\Omega(x^+_h,x^-_h) = 0$ now gives
\begin{equation}
 x^+_h\biggl(x^-_h + \frac{m}{\lambda^3x^+_0}\biggr) = -\frac{\kappa}{2\lambda^2}
 \,,
\end{equation}
and this equation defines a curve which is partially contained in the region $\{x^+>0, x^-<0\}$ and across which the sign of $(\nabla\phi)^2$ changes.

\begin{figure}[htb]
 \centering
 \includegraphics[scale=1]{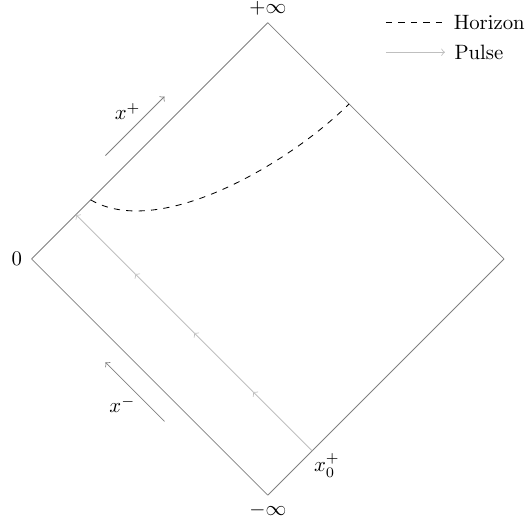}
 \caption{Boulware spacetime for non-physical particles perturbed by a shock wave of classical matter at $x^+=x^+_0$ which creates an apparent horizon, but the solution remains singularity free.}
 \label{BNP_DynamicSpacetime}
\end{figure}

Therefore the Boulware solution for non-physical particles is always singularity free and sending in a shock wave of matter creates an apparent horizon which is absent in the static solution. This spacetime is represented on figure \ref{BNP_DynamicSpacetime}.
\subsection{Boulware state: physical particles}
Let us now study the Boulware state ($P=-\frac{1}{4}$) for physical particles ($\kappa > 0$) which can also be a plausible scenario.
\subsubsection{Static solution}
The static solution is
\begin{equation}
 \Omega = \kappa\phi + e^{-2\phi} = -\lambda^2x^+x^- - \frac{\kappa}{2}\ln(-\lambda^2x^+x^-) + \frac{M}{\lambda}
 \,, \;
 \kappa > 0
 \,.
 \label{RST_OmegaStaticSolutionB_P}
\end{equation}
\paragraph{Singularity:} the noticeable difference with the non-physical case ($\kappa < 0$) is that now $\Omega' = \kappa - 2e^{-2\phi}$ vanishes at $\phi = \phi_{cr} = -\frac{1}{2}\ln\frac{\kappa}{2}$ where $\Omega$ takes the value $\Omega_{cr}$ defined by \eqref{RST_OmegaCritical}. This means there is a possible curvature singularity on the curve $(x^+_s,x^-_s)$ defined by
\begin{equation}
 \frac{\kappa}{2}\biggl(1 - \ln\frac{\kappa}{2}\biggr) = - \lambda^2x_s^+x_s^- - \ln(-\lambda^2x_s^+x_s^-) + \frac{M}{\lambda}
 \,.
\end{equation}
If we define
\begin{equation}
 Z_s \equiv - \frac{2\lambda^2}{\kappa}x_s^+x_s^-
 \,,
\end{equation}
this equation can be rewritten as
\begin{equation}
 W(Z_s) \equiv Z_s - \ln Z_s = 1 - \frac{2M}{\lambda\kappa}
 \,.
\end{equation}
For $Z_s>0$ the function $W$ has a minimum at $Z_s=1$ which is $W(1) =1$ so for $M > 0$ this equation has no solution in the region $Z_s > 0$ and this singularity is not in the region $\{x^+ > 0, x^-<0\}$.

Therefore we need to look at what happens where $x^+x^-=0$. As in the $\kappa < 0$ case, let us study the limit $x^- \rightarrow 0$ for a fixed value of $x^+$. In this limit $\Omega \rightarrow +\infty$, which corresponds to $\phi \rightarrow \pm \infty$, but in order to get to $\phi \rightarrow +\infty$ one would have to pass through the singularity at $\phi = \phi_{cr}$ (as $\phi \rightarrow -\infty$ at spatial infinity). Thus the relevant limit is $\phi \rightarrow -\infty$ as $x^-\rightarrow 0$ (at a fixed value of $x^+$). Let us compute the curvature in this limit. First we have
\begin{equation}
 \Omega \sim e^{-2\phi} \sim -\frac{\kappa}{2}\ln(-\lambda^2x^+x^-)
 \,, \quad
 \Omega' \sim -2e^{-2\phi}
 \,, \quad
 \Omega'' = 4e^{-2\phi}
 \,,
\end{equation}
and
\begin{equation}
 \partial_\pm\Omega = -\lambda^2x^\mp -\frac{\kappa}{2x^\pm} \sim -\frac{\kappa}{2x^\pm}
 \,, \quad
 \partial_+\partial_-\Omega = -\lambda^2
 \,,
\end{equation}
so in this limit the curvature \eqref{RST_CurvatureOmega} is
\begin{equation}
 R \sim 4\biggl(\lambda^2 + \frac{\Omega''}{\Omega'^2}\partial_+\Omega\partial_-\Omega\biggr)
 \,,
\end{equation}
with
\begin{equation}
 \frac{\Omega''}{\Omega'^2}\partial_+\partial_-\Omega
 \sim
 \frac{\kappa^2}{4e^{-2\phi}x^+x^-} \sim \frac{-\kappa}{2x^+x^-\ln(-\lambda^2x^+x^-)} \rightarrow -\infty
 \,.
\end{equation}
so there is a null singularity on the axes $x^\pm = 0$.

\paragraph{Geodesic completeness:}we can check whether null geodesics are complete when approaching this singularity by considering the integral
\begin{equation}
 \Delta\chi = \int_{x^-_0}^0\mathrm{d}x^-\,e^{2\phi(x^+_0,x^-)}
 \,.
\end{equation}
This time we have $e^{2\phi(x^+_0,x^-)} \sim \frac{-2}{\kappa\ln(-\lambda^2x^+_0x^-)} \rightarrow 0$ when $x^-\rightarrow 0$ so this integral is convergent, meaning null geodesics are incomplete. In static coordinates a careful analysis shows that timelike geodesics are also incomplete at this singularity.

\paragraph{Horizon vs type I wormhole:} the conditions $\partial_\pm\Omega(x^+_h,x^-_h) = 0$ gives
\begin{equation}
 x^+_hx^-_h = -\frac{\kappa}{2\lambda^2} < 0
 \,,
\end{equation}
which defines a curve $(x^+_h,x^-_h)$ contained in the region $\{x^+ > 0, x^-<0\}$, but the sign of $(\nabla\phi)^2$ does not change when crossing this curve, meaning that this is not an apparent horizon but rather the throat of a Type I wormhole. On one side there is the null singularity and on the other an asymptotically flat spacetime. This is what we had already found in \cite{potaux_quantum_2022}.

\begin{figure}[htb]
 \centering
 \includegraphics[scale=1]{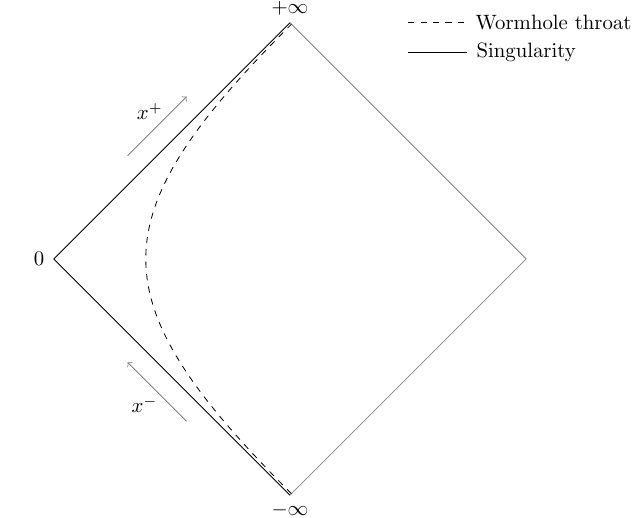}
 \caption{Static Boulware spacetime for physical particles, there is a null singularity at $x^+x^-=0$ on the other side of a type I wormhole throat. Geodesics are incomplete when approaching the singularity.}
 \label{BP_StaticSpacetime}
\end{figure}

The static Boulware solution for physical particles thus contains a singularity located behind the throat of a wormhole, it is represented on figure \ref{BP_StaticSpacetime}.

\subsubsection{Dynamical solution}
The corresponding dynamical solution is given by
\begin{equation}
 \Omega = \kappa \phi + e^{-2\phi}
 = -\lambda^2x^+x^- - \frac{\kappa}{2}\ln(-\lambda^2x^+x^-)
 + \frac{M}{\lambda} - \frac{m}{\lambda x^+_0}(x^+-x^+_0)\theta(x^+-x^+_0)
 \,, \quad
 \kappa > 0
 \,,
\end{equation}
which for $x^+ > x^+_0$ becomes
\begin{equation}
 \Omega = -\lambda^2x^+\tilde{x}^- - \frac{\kappa}{2}\ln(-\lambda^2x^+x^-) + \frac{M+m}{\lambda}
 \,, \quad
 \tilde{x}^- = x^- + \frac{m}{\lambda^3x^+_0}
 \,.
\end{equation}
\paragraph{Horizons:} we notice that there are now two distinct apparent horizons as
\begin{equation}
 \left\{
 \begin{aligned}
  \partial_-\Omega & = 0 \Leftrightarrow x^+x^- = -\frac{\kappa}{2\lambda^2}
  \,, & (\mathcal{H}_1)
  \\
  \partial_+\Omega & = 0 \Leftrightarrow x^+\biggl(x^-+\frac{m}{\lambda^3x^+_0}\biggr) = -\frac{\kappa}{2\lambda^2}
  \,, & (\mathcal{H}_2)
 \end{aligned}
 \right.
 \label{BP_Dynamic_Horizons}
\end{equation}
and the sign of $(\nabla\phi)^2$ changes sign when crossing these two curves. The apparent horizon $\mathcal{H}_1$ is the continuation of what was the wormhole throat before the shock wave, while $\mathcal{H}_2$ is a new apparent horizon.

\paragraph{Singularities:} as in the $\kappa > 0$ case the limit $x^+x^- \rightarrow 0$ gives the same behavior as before the shock wave so there is still a singularity there. However we have to check whether the singularity at $\phi = \phi_{cr}$ has moved into the region $\{x^+ > 0, x^-<0\}$. It is on the curve $(x_s^+,x_s^-)$ defined by
\begin{equation}
 F(x^+_s,x^-_s) = -\lambda^2x_s^+\biggl(x_s^- + \frac{m}{\lambda^3x^+_0}\biggr) - \frac{\kappa}{2}\ln(-\lambda^2x_s^+x_s^-) + \frac{M+m}{\lambda} - \frac{\kappa}{2}\biggl(1 - \ln\frac{\kappa}{2}\biggr) = 0
 \,,
\end{equation}
which can be rewritten as
\begin{equation}
 W(Z_s) = Z_s - \ln Z_s = 1 - \frac{2(M+m)}{\lambda\kappa} + \frac{2m}{\lambda\kappa x^+_0}x_s^+
 \,, \quad
 Z_s = -\frac{2\lambda^2}{\kappa}x_s^+x_s^-
 \,.
\end{equation}
For each value of $x_s^+ > x^+_m \equiv \Bigl(1+\frac{M}{m}\Bigr)x^+_0$ the right-hand-side of this equation is greater than $1$ so there are two solutions, one on the branch $Z_s < 1$, located behind the horizon $\mathcal{H}_1$ (which can be defined by $Z=1$), and one on the branch $Z_s > 1$, located before this horizon.

\begin{figure}[htb]
 \centering
 \includegraphics[scale=1]{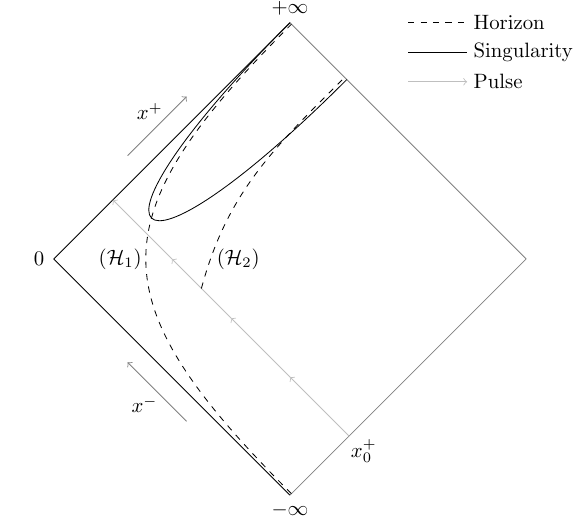}
 \caption{Boulware spacetime for physical particles perturbed by a shock wave of classical matter at $x^+=x^+_0$. What was a wormhole throat before the shock becomes an apparent horizon $\mathcal{H}_1$ and a singularity appears. The upper branch of the singularity is timelike  and the lower one is spacelike behind the horizon $\mathcal{H}_2$ and it becomes timelike after they collide at a finite point.}
 \label{BP_DynamicSpacetime}
\end{figure}

Therefore a singularity appears at $x^+=x^+_m$, and to see whether it is spacelike or timelike we study the sign of
\begin{equation}
 S(x^+_s,x^-_s) \equiv \frac{\partial_+F}{\partial_-F} = \frac{x^-_s}{x^+_s}\biggl(1+\frac{\frac{2m}{\lambda\kappa x^+_0}x^+_s}{1+\frac{2\lambda^2}{\kappa}x^+_sx^-_s}\biggr)
 = \frac{x^-_s}{x^+_s}\biggl(1+\frac{\frac{2m}{\lambda\kappa x^+_0}x^+_s}{1-Z_s}\biggr)
 \,.
\end{equation}
On the branch $Z_s < 1$ this quantity is always negative (note that $\frac{x^-_s}{x^+_s}<0$) so that the singularity is timelike. For the other branch $Z_s > 1$ we have, in the limit $Z_s \rightarrow 1^+$, $S(x^+_s,x^-_s) \rightarrow +\infty$ and in the limit $Z_s \rightarrow +\infty$ we get $S(x^+_s,x^-_s) \sim \frac{x^-_s}{x^+_s} < 0$. Therefore on this branch the singularity is spacelike at first and then it becomes timelike. The point at which this happens is such that $S(x^+_s,x^-_s) = 0$ which gives
\begin{equation}
 1 + \frac{2\lambda^2}{\kappa}x^+_s\biggl(x^-_s + \frac{m}{\lambda^3x^+_0}\biggr) = 0
 \,.
\end{equation}
This is actually the equation satisfied by the horizon $\mathcal{H}_2$ \eqref{BP_Dynamic_Horizons}, which means that the singularity collides with this horizon and becomes timelike at a finite point $(x^+_c,x^-_c)$, which is given by
\begin{equation}
 \left\{
 \begin{aligned}
  x^+_c & = \frac{\lambda\kappa x^+_0}{2m}(e^{2m/\lambda\kappa}-1)
  \,,
  \\
  x^-_c & = -\frac{m}{\lambda^3x^+_0}\frac{1}{1-e^{-2m/\lambda\kappa}}
  \,.
 \end{aligned}
 \right.
\end{equation}
Thus the singularity is naked for $x^+ > x^+_c$. Note that both the singularity and $\mathcal{H}_2$ have the same asymptote $x^- = -\frac{m}{\lambda^3x^+_0}$ when $x^+ \rightarrow +\infty$. This solution is represented on figure \ref{BP_DynamicSpacetime}.

\section{A hybrid quantum state}
\label{SectionHybridRST}

Now that we have studied in detail the various solutions corresponding to the Hartle-Hawking and Boulware states when all the particles are in the same state, let us focus on a hybrid quantum state with two auxiliary fields $\psi_1$ and $\psi_2$, with central charges $\kappa_1$ and $\kappa_2$ respectively. We will assume that $\psi_1$ describes physical particles in the Hartle-Hawking state, so that $\kappa_1 > 0$ and $P_1=0$, and that $\psi_2$ describes non-physical particles in the Boulware state, meaning that $\kappa_2 < 0$ and $P_2 = -\frac{1}{4}$. This choice is motivated by the fact that an observer at infinity should only detect the radiation of physical particles. The static solution \eqref{RST_OmegaStaticSolution} then becomes
\begin{equation}
 \Omega = \kappa\phi + e^{-2\phi} = -\lambda^2x^+x^- - \frac{\kappa_2}{2}\ln(-\lambda^2x^+x^-) + \frac{M}{\lambda}
 \,, \quad
 \kappa = \kappa_1 + \kappa_2
 \,,
\end{equation}
and the dynamical solution for $x^+ > x^+_0$ \eqref{RST_OmegaDynamicalSolution}
\begin{equation}
 \Omega= \kappa\phi + e^{-2\phi} = -\lambda^2x^+\biggl(x^-+\frac{m}{\lambda^3x^+_0}\biggr)
 - \frac{\kappa_2}{2}\ln(-\lambda^2x^+x^-)
 + \frac{M+m}{\lambda}
 \,.
\end{equation}
We will study the three cases $\kappa = 0$, $\kappa > 0$ and $\kappa < 0$ separately in the following.

\subsection{Zero $\kappa$ case}
Let us assume that $\kappa = \kappa_1 + \kappa_2 = 0$, meaning that the contribution of non-physical particles exactly compensates the physical particle's. In this case $\Omega = e^{-2\phi} > 0$.

\subsubsection{Static solution}
\begin{figure}[hbt]
 \centering
 \includegraphics[scale=1]{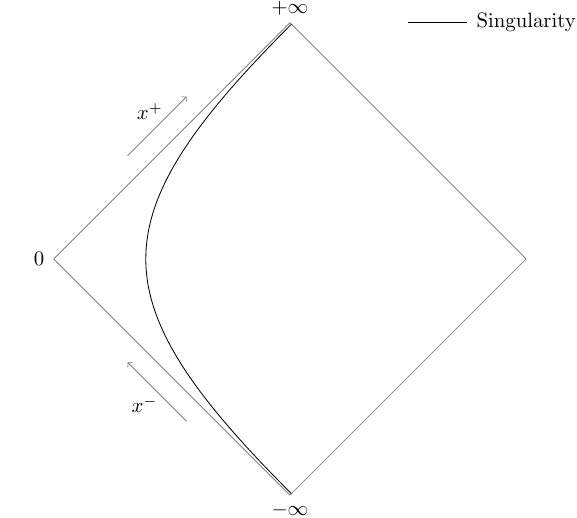}
 \caption{Static spacetime for the Hybrid RST model with $\kappa = 0$ and $\kappa >0$. The singularity is timelike and naked, and geodesics are incomplete when approaching it.}
 \label{Kappa=0_StaticSpacetime}
\end{figure}

The static solution takes the form
\begin{equation}
 e^{-2\phi} = -\lambda^2x^+x^- - \frac{\kappa_2}{2}\ln(-\lambda^2x^+x^-) + \frac{M}{\lambda}
 \,,
\end{equation}
which can be rewritten as
\begin{equation}
 e^{-2\phi} = h(Z) + \frac{M}{\lambda}
 \,, \quad
 Z = -\lambda^2x^+x^-
 \,, \quad
 h(Z) \equiv Z - \frac{\kappa_2}{2}\ln Z
 \,. 
\end{equation}
Note that $Z>0$ on the region $\{x^+>0,x^-<0\}$. Since $e^{-2\phi} > 0$ only the region $h(Z) > -\frac{M}{\lambda}$ is available, which corresponds to  $Z>Z_s$ where $Z_s$ is the only solution of $h(Z)=-\frac{M}{\lambda}$ ($\kappa_2 < 0$ so the function $h$ is increasing on $]0,+\infty[$). Therefore the available spacetime is the region $x^+x^- < -\frac{Z_s}{\lambda^2}$.
\paragraph{Horizon:} we have
\begin{equation}
 \partial_\pm\Omega(x^+_h,x^-_h) = 0 \Leftrightarrow
 x^+_hx^-_h =  -\frac{\kappa_2}{2\lambda^2} > 0
 \,,
\end{equation}
so there is no horizon in the spacetime.
\paragraph{Singularity:} the curvature \eqref{RST_CurvatureOmega} can be computed explicitly in terms of $(x^+,x^-)$ to get
\begin{equation}
 R = 4\biggl\{\lambda^2 + \frac{e^{2\phi}}{x^+x^-}\biggl(\lambda^2x^+x^- + \frac{\kappa_2}{2}\biggr)^2\biggr\}
 \,,
\end{equation}
and it diverges when $x^+x^- \rightarrow -\frac{Z_s}{\lambda^2}$ since $\Omega \rightarrow 0$ and $\phi \rightarrow + \infty$ in this limit. Therefore the border $x^+x^- = -\frac{Z_s}{\lambda^2}$ corresponds to a curvature singularity. Applying the criteria from Appendix \ref{Appendix_TimelikeSpacelike} we get that this singularity is timelike.
\paragraph{Geodesic completeness:}let us check whether geodesics are complete when approaching this singularity. For null geodesics we look at the integral
\begin{equation}
 \Delta\chi = \int_{x^-_0}^{-Z_s/\lambda^2x^+_0}\d x^-\,e^{2\phi(x^+_0,x^-)}
 \,.
\end{equation}
As $x^- \rightarrow -\frac{Z_0}{\lambda^2x^+_0}$ we have $e^{2\phi} = \Omega^{-1} \sim [f'(Z_0)(Z-Z_0)]^{-1}$ so this integral diverges and null geodesics are complete in this direction. However, using static coordinates and the fact that the function $Z$ is finite at the singularity one can show that timelike geodesics are incomplete there. Therefore this spacetime contains a naked singularity and is geodesically incomplete. It is represented on figure \ref{Kappa=0_StaticSpacetime}.

\subsubsection{Dynamical solution}
\begin{figure}[hbt]
 \centering
 \includegraphics[scale=1]{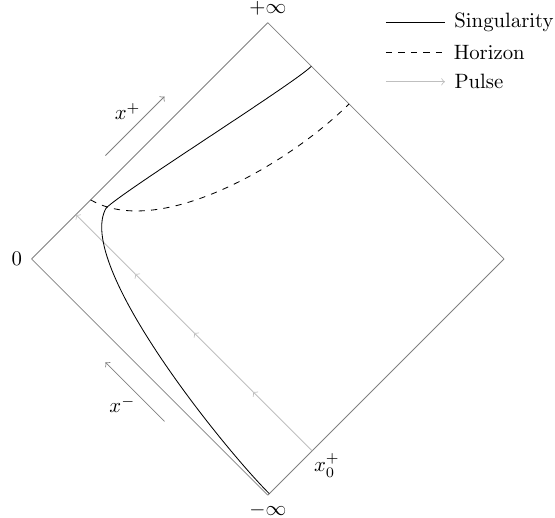}
 \caption{Dynamical spacetime for the Hybrid RST model with $\kappa = 0$. At first the singularity is naked and timelike but it goes behind the apparent horizon at a finite point and becomes spacelike. This picture also applies to the case $\kappa > 0$.}
 \label{Kappa=0_DynamicSpacetime}
\end{figure}

The dynamical solution is
\begin{equation}
 e^{-2\phi} = -\lambda^2x^+x^-
 - \frac{\kappa_2}{2}\ln(-\lambda^2x^+x^-)
 + \frac{M}{\lambda} - \frac{m}{\lambda x^+_0}(x^+-x^+_0)\theta(x^+-x^+_0)
 \,,
\end{equation}
which, for $x^+ > x^+_0$, becomes
\begin{equation}
 e^{-2\phi} = h(Z) - \frac{m}{\lambda x^+_0}x^+ + \frac{M+m}{\lambda}
 \,, \quad
 Z = -\lambda^2x^+x^-
 \,, \quad
 h(Z) = Z - \frac{\kappa_2}{2}\ln Z
 \,.
\end{equation}
Here the available spacetime is the region where $h(Z) > \frac{m}{\lambda x^+_0}x^+ - \frac{M+m}{\lambda}$. For a fixed $x^+$ this corresponds to $Z > Z_s(x^+)$ where $Z_s(x^+)$ is the only solution of $h(Z) = \frac{m}{\lambda x^+_0}x^+ - \frac{M+m}{\lambda}$.
\paragraph{Horizon:}contrary to the static case, here an apparent horizon appears as
\begin{equation}
 \partial_+\Omega (x^+_h, x^-_h) = 0 \Leftrightarrow
 x^+_h\biggl(x^-_h + \frac{m}{\lambda^3x^+_0}\biggr) = -\frac{\kappa_2}{2\lambda^2}
 \,,
\end{equation}
and this equation has a solution in the studied region for $x+ \geq -\frac{\lambda\kappa_2x^+_0}{2m}$. When $x^+_h \rightarrow +\infty$ we have $x^-_h \rightarrow -\frac{m}{\lambda^3x^+_0}$ which is therefore the asymptote of the horizon.
\paragraph{Singularity:}compared to the static case, the trajectory of the curve $(x^+_s,x^-_s)$ defined by $\Omega(x^-_s,x^-_s) = 0$ is modified as this gives
\begin{equation}
 x^+_sx^-_s = -\frac{Z_s(x^+_s)}{\lambda^2}
 \,.
\end{equation}
As in the static case it corresponds to a singularity as the curvature
\begin{equation}
 R = 4\biggl\{\lambda^2 + \frac{e^{2\phi}}{x^+x^-}\biggl[\lambda^2x^+\biggl(x^- + \frac{m}{\lambda^3x^+_0}\biggr) + \frac{\kappa_2}{2}\biggr]\biggl(\lambda^2x^+x^- + \frac{\kappa_2}{2}\biggr)\biggr\}
 \,,
\end{equation}
diverges when $\phi \rightarrow +\infty$ and $x^\pm$ go to finite values. On this singularity curve we also have $x^-_s \rightarrow -\frac{m}{\lambda^3x^+_0}$ when $x^+_s\rightarrow \infty$ so the apparent horizon and the singularity have the same asymptote. Let us study their relative behavior. We have
\begin{equation}
 \left\{
 \begin{aligned}
  x^-_h(x^+) & = \frac{\kappa_1}{2\lambda^2x^+} - \frac{m}{\lambda^3x^+_0}
  \,, \\
  x^-_s(x^+) & = \frac{\kappa_1}{2\lambda^2x^+}\ln(-\lambda^2x^+x^-_s) + \frac{M+m}{\lambda^2x^+} - \frac{m}{\lambda^3x^+_0}
  \,,
 \end{aligned}
 \right.
\end{equation}
so
\begin{equation}
 \tilde{x}^-_s(x^+) - \tilde{x}^-_h(x^+) = \frac{\kappa_1}{2\lambda^2x^+}\big\{\ln(-\lambda^2x^+x^-_s)-1\big\} + \frac{M+m}{\lambda^2x^+}
 \,,
\end{equation}
and this is positive for sufficiently large values of $x^+$ so that $x^-_s > x^-_h$. The point of intersection between these two curves is at $(x^+_{hs},x^-_{hs})$ such that $\partial_+\Omega(x^+_{hs},x^-_{hs}) = 0 = \Omega(x^+_{hs},x^-_{hs})$ which leads to
\begin{equation}
 \left\{
 \begin{aligned}
  x^+_{hs} & = \frac{\lambda x^+_0}{m}\biggl(\frac{\kappa_1}{2} + e^{1-2(M+m)/\kappa_1}\biggr)
  \,, \\
  x^-_{hs} & = \frac{-m}{\lambda^3x^+_0\biggl(\frac{\kappa_1}{2}e^{-1+2(M+m)/\kappa_1} + 1\biggr)} 
  \,.
 \end{aligned}
 \right.
\end{equation}
Besides, as $\Omega(x^+_s,x^-_s) = 0$, we have
\begin{equation}
 \frac{\d x^-_s}{\d x^+} = - \frac{\partial_+\Omega}{\partial_-\Omega}
\end{equation}
so the maximum of the curve $x_s^-(x^+)$ corresponds to the intersection with the apparent horizon at $(x^+_{hs},x^-_{hs})$. Since $\partial_-\Omega = -\lambda^2x^- - \frac{\kappa_2}{2x^+} > 0$ the nature of the singularity is determined by the sign of $\partial_+\Omega$, \textit{i.e.}\ by the position of the singularity relative to the horizon (defined by $\partial_+\Omega = 0$). At first it is outside of the horizon, where $\partial_+\Omega < 0$, so it is timelike. After the intersection with the horizon it is located inside it, where $\partial_+\Omega > 0$ and it becomes spacelike. This spacetime is represented on figure \ref{Kappa=0_DynamicSpacetime}.

\subsection{Positive $\kappa$ case}
Let us now consider the case where $\kappa > 0$. Then $\Omega$ has a minimum at $\phi = \phi_{cr} = -\frac{1}{2}\ln\frac{\kappa}{2}$ and takes the value $\Omega_{cr} = \frac{\kappa}{2}\Bigl(1-\ln\frac{\kappa}{2}\Bigr)$. Therefore the situation is essentially the same as in the previous case ($\kappa = 0$), the only difference being that the minimal value of $\Omega$ is now reached at a finite value of $\phi$. As we will see the global structure will thus be almost the same.

\subsubsection{Static solution}
The static solution can be written as
\begin{equation}
 \Omega = \kappa\phi + e^{-2\phi} = h(Z) + \frac{M}{\lambda}
 \,, \quad
 Z = -\lambda^2x^+x^-
 \,, \quad
 h(Z) = Z -\frac{\kappa_2}{2}\ln Z
 \,,
\end{equation}
and only the region $f(Z) > \Omega_{cr}-\frac{M}{\lambda}$ is available. This corresponds to  $Z>Z_s$ where $Z_s$ is the only solution of $h(Z)=\Omega_{cr}-\frac{M}{\lambda}$. Therefore the available spacetime is the region $x^+x^- < -\frac{Z_s}{\lambda^2}$.
\paragraph{Horizon:} as in the $\kappa = 0$ case the apparent horizon condition gives
\begin{equation}
 \partial_\pm\Omega (x^+_h,x^-_h) = 0 \Leftrightarrow
 x^+_hx^-_h =  -\frac{\kappa_2}{2\lambda^2} > 0
 \,,
\end{equation}
so there is no apparent horizon in the spacetime.

\paragraph{Singularity:}the curvature is
\begin{equation}
 R = \frac{8e^{-2\phi}}{\Omega'}\biggl\{-\lambda^2 - \frac{4e^{-2\phi}}{\Omega'^2x^+x^-}\biggl(\lambda^2x^+x^- + \frac{\kappa_2}{2}\biggr)^2\biggr\}
 \,,
\end{equation}
and it diverges when $x^+x^- \rightarrow -\frac{Z_s}{\lambda^2}$ since $\phi \rightarrow \phi_{cr}$ and $\Omega'\rightarrow 0$ in this limit which means that the border $x^+x^- = -\frac{Z_s}{\lambda^2}$ corresponds to a timelike singularity.

\paragraph{Geodesic completeness:} we can see that this null geodesics are incomplete when approaching this singularity by looking at the integral
\begin{equation}
 \Delta\chi = \int_{x^-_0}^{-Z_0/\lambda^2x^+_0}\d x^-\,e^{2\phi(x^+_0,x^-)}
 \,,
\end{equation}
which is convergent since $\phi(x^+_0, x^-) \rightarrow \phi_{cr}$ when $x^- \rightarrow -\frac{Z_0}{\lambda^2x^+_0}$. As in the $\kappa = 0$ case timelike geodesics are also incomplete there, as $Z$ goes to a finite value. The spacetime therefore has the same structure as in the $\kappa =0$ case, with the only difference being that timelike geodesics are incomplete near the singularity.
\subsubsection{Dynamical solution}
As in the static case the dynamical spacetime for $\kappa > 0$ has the same structure as for $\kappa =0$.
\subsection{Negative $\kappa$ case}
Finally let us consider the case $\kappa < 0$, meaning the number of non-physical particles exceeds the number of physical particles. The properties of this solution were already described in \cite{potaux_space-time_2022} so we will recall the most interesting ones.
\subsubsection{Static solution}
Here the static solution is given by
\begin{equation}
 \Omega = \kappa\phi + e^{-2\phi} = -\lambda^2x^+x^- - \frac{\kappa_2}{2}\ln(-\lambda^2x^+x^-) + \frac{M}{\lambda}
 \,, \quad
 \kappa < 0
 \,.
 \label{HRST_KappaNegative_StaticSolution}
\end{equation}

\paragraph{Singularity:} first, since $\Omega' = \kappa - 2e^{-2\phi} < 0$ never vanishes, there is no associated singularity. Then it appears that $\Omega$ diverges when $x^\pm \rightarrow 0$ so let us study the curvature as $x^- \rightarrow 0$ for a fixed value of $x^+$ (by symmetry the result will also be valid for $x^+\rightarrow 0$). Since $\kappa_2 < 0$, $\Omega \rightarrow -\infty$ so $\phi \rightarrow +\infty$ and we get
\begin{equation}
 \Omega \sim -\frac{\kappa_2}{2}\ln(-\lambda^2 x^+x^-) \sim \kappa\phi
 \,, \quad
 \Omega' \sim \kappa
 \,, \quad
 \Omega'' = 4e^{-2\phi}
 \,,
\end{equation}
and
\begin{equation}
 \partial_+\Omega \rightarrow -\frac{\kappa_2}{2x^+}
 \,, \quad
 \partial_-\Omega \sim -\frac{\kappa_2}{2x^-} \rightarrow \pm \infty
 \,, \quad
 \partial_+\partial_-\Omega = -\lambda^2
 \,.
\end{equation}
The first curvature term in \eqref{RST_CurvatureOmega} goes to zero as $e^{-2\phi}$ so we have to compute the behavior of the second term
\begin{equation}
 R - O(e^{-2\phi}) = -\frac{8e^{-2\phi}\Omega''}{\Omega'^3}\partial_+\Omega\partial_-\Omega \sim -\frac{8\kappa_2^2}{\kappa^3}\frac{e^{-4\phi}}{x^+x^-}
 \,.
\end{equation}
To do this we use the fact that
\begin{equation}
 -\kappa\phi - \frac{\kappa_2}{2}\ln(-\lambda^2x^+x^-) + \frac{M}{\lambda} = \lambda^2x^+x^- + e^{-2\phi} \rightarrow 0
 \,,
\end{equation}
so
\begin{equation}
 -\lambda^2x^+x^- \sim e^{2M/\lambda\kappa_2}e^{-2\kappa\phi/\kappa_2}
 \,,
 \label{HRST_Kappa<0_Simx+x-}
\end{equation}
and
\begin{equation}
 R - O(e^{-2\phi}) \sim \frac{8\lambda^2\kappa_2^2}{k^3}e^{-2M/\lambda \kappa_2}e^{-2(2-\kappa/\kappa_2)\phi}
 \,.
\end{equation}
Since $0> \kappa > \kappa_2$ so $\frac{\kappa}{\kappa_2} < 1$ and $2-\frac{\kappa}{\kappa_2} > 0$ which means that $R\rightarrow 0$ so there is no curvature singularity when $x^\pm \rightarrow 0$. Therefore the region $\{x^+>0,x^-<0\}$ is singularity free.

\begin{figure}[hbt]
 \centering
 \includegraphics[scale=1]{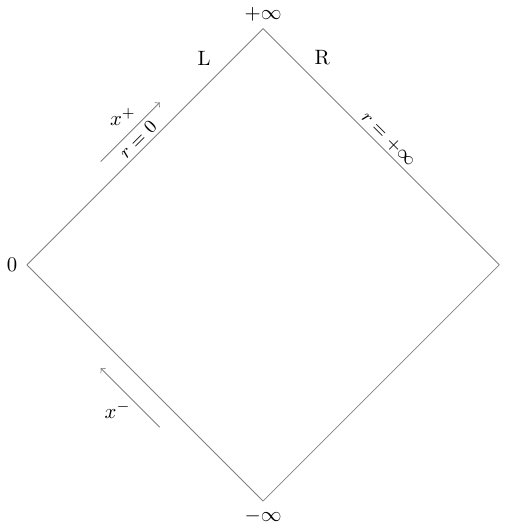}
 \caption{Static spacetime for the Hybrid RST model with $\kappa < 0$. There is no singularity and no horizon and it is geodesically complete. We have used the variable $r=e^{-\phi}$ which can be seen as a radius.}
 \label{KappaNegative_StaticSpacetime}
\end{figure}

\paragraph{Geodesic completeness:} let us see if null geodesics are complete when $x^+x^- \rightarrow 0$ by considering the integral
\begin{equation}
 \Delta\chi = \int_{x^-_0}^{0}\d x^-\,e^{2\phi(x^+_0,x^-)}
 \,,
\end{equation}
According to \eqref{HRST_Kappa<0_Simx+x-} when $x^-\rightarrow 0$ we have
\begin{equation}
 e^{2\phi} \sim e^{2M/\lambda \kappa}(-\lambda^2x^+x^-)^{-\kappa_2/\kappa}
 \,,
\end{equation}
where $\kappa_2/\kappa > 1$ so that the integral is divergent and the geodesic is complete. For timelike geodesics we use static coordinates and \eqref{HRST_Kappa<0_Simx+x-} which can be rewritten as
\begin{equation}
 Z(\phi) = O(e^{-2\kappa\phi/\kappa_2})
 \,.
\end{equation}
Then we have to compute the integral \eqref{DeltaLambdaConformalTimelike} where, when $\phi \rightarrow +\infty$,
\begin{equation}
 \sqrt{\frac{h^2(\phi)}{E^2-g(\phi)}} = O(e^{2(1-\kappa/\kappa_2)\phi}) \rightarrow +\infty
 \,,
\end{equation}
since $0< \kappa/\kappa_2 < 1$. Thus the integral is divergent and timelike geodesics are also complete when $x^+x^- \rightarrow 0$.
Therefore the region $\{x^+>0,x^-<0\}$ is geodesically complete. We can also check this by performing a change of coordinates. Let us define
\begin{equation}
 y^\pm \equiv \pm\lambda^{-\kappa_2/\kappa}e^{M/\lambda \kappa}\frac{\kappa}{\kappa_1}\frac{1}{(\pm x^\pm)^{-\kappa_1/\kappa}}
 \,,
\end{equation}
such that
\begin{equation}
 \d s^2 \sim -\d y^+\d y^-
 \,,
\end{equation}
when $x^\pm \rightarrow 0^\pm$, so the metric is regular in this limit which corresponds to $y^\pm \rightarrow \mp \infty$.
In these coordinates the metric is given by
\begin{equation}
 \d s^2 = -e^{2M/\lambda k}e^{-2\phi}(-\lambda^2x^+x^-)^{-\kappa_2/k}\d y^+ \d y^-
 \,.
\end{equation}
In the limit $x^\pm \rightarrow 0^\pm$ we have $y^\pm \rightarrow \mp\infty$ so the geodesics are complete in these directions.
\paragraph{Horizon:}the apparent horizon condition $\partial_\pm\Omega=0(x^+_h,x^-_h)$ gives
\begin{equation}
 x^+_hx^-_h = -\frac{\kappa_2}{2\lambda^2} > 0
 \,,
\end{equation}
so there is no apparent horizon in this spacetime. It is represented on figure \ref{KappaNegative_StaticSpacetime}, where we used the variable $r=e^{-\phi}$ which can be seen as a radius.

\paragraph{Type II wormhole structure (black hole mimicker):} as explained in \cite{potaux_space-time_2022} this solution is a type II wormhole, \textit{i.e.}\ a black hole mimicker as suggested in \cite{damour_wormholes_2007} with a throat where time flows slowly compared to an external observer.

\subsubsection{Dynamical solution}
Let us now consider the dynamical solution for this $\kappa < 0$ case. For $x^+ > x^+_0$ we get
\begin{equation}
 \Omega = \kappa\phi + e^{-2\phi} = -\lambda^2x^+\biggl(x^-+\frac{m}{\lambda^3x^+_0}\biggr) - \frac{\kappa_2}{2}\ln(-\lambda^2x^+x^-) + \frac{M+m}{\lambda}
 \,.
 \label{KappaNegative_DynamicalSolution}
\end{equation}
\paragraph{Horizon:}an apparent horizon appears because of the shock wave since the apparent horizon condition now gives
\begin{equation}
 \partial_+\Omega(x^+_h,x^-_h) = 0 \Leftrightarrow
 x^+_h\biggl(x^-_h+\frac{m}{\lambda^3x^+_0}\biggr) = -\frac{\kappa_2}{2\lambda^2}
 \quad (\mathcal{H})
 \,,
\end{equation}
and this is the equation of a curve $\mathcal{H}$ partially contained in the region $\{x^+>0,x^-<0\}$, which correspond to an apparent horizon. Its asymptote when $x^+ \rightarrow +\infty$ is $x^-_h = -\frac{m}{\lambda^3x^+_0}$.
\paragraph{Singularity:} the limit $x^- \rightarrow 0$ is essentially the same as in the static case and a similar computation shows that the curvature goes to zero there. We can also check what happens for the curvature at future null infinity, \textit{i.e.}\ when $x^+ \rightarrow +\infty$ for a fixed value of $x^-$. We have to distinguish according to the sign of $\tilde{x}^- = x^- -x^-_h$, that is to say according to the position relative to the asymptote of the apparent horizon $\mathcal{H}$.
\begin{enumerate}
 \item Inside the horizon $x^- > x^-_h$: here $\Omega \rightarrow -\infty$ so $\phi \rightarrow +\infty$ and we get that $R \rightarrow 0$.
 \item Outside the horizon $x^- < x^-_h$: we have $\Omega \rightarrow +\infty$ so $\phi \rightarrow -\infty$ and it is easy to check that $R \rightarrow 0$.
 \item Along the horizon $x^- = x^-_h$: here we also have $\Omega \rightarrow +\infty$ and $\phi \rightarrow -\infty$ but computation of the limit shows that $R \rightarrow 4\lambda^2$, which means that there seems to be a curvature discontinuity along the future null infinity at point $\mathcal{C}$, which is defined by $x^+ \rightarrow +\infty$ for $x^- = -\frac{m}{\lambda^3x^+_0}$.
\end{enumerate}
This curvature discontinuity deserves a bit more analysis. Let us study the behavior of timelike geodesics in the limit $x^+ \rightarrow +\infty$ and outside of the horizon ($x^- < x^-_h$). The geodesic equation for $x^+$ is
\begin{equation}
 \frac{\d^2x^+}{\d\tau^2} + 2\partial_+\phi\biggl(\frac{\d x^+}{\d\tau}\biggr)^2 = 0
 \,,
\end{equation}
where $\tau$ denotes the proper time along the geodesic. When $x^+$ goes to infinity we have $\partial_+\phi \sim -\frac{1}{2x^+}$ and we can solve the geodesic equation perturbatively to get that
\begin{equation}
 x^+ \simeq e^{\alpha\tau}
 \,, \quad
 \alpha > 0
 \,.
\end{equation}
Inserting this into the timelike geodesic condition
\begin{equation}
 \biggl(\frac{\d s}{\d\tau}\biggr)^2 = -e^{2\phi}\frac{\d x^+}{\d\tau}\frac{\d x^-}{\d\tau} = -1
 \,,
\end{equation}
and solving for $\tilde{x}^- = x^- - x^-_h$ one gets that
\begin{equation}
 \tilde{x}^- \simeq -e^{-\frac{\lambda^2}{\alpha}\tau}
 \,.
\end{equation}
Note that we have $x^+\tilde{x^-} \simeq -e^{\alpha\bigl(1-\frac{\lambda^2}{\alpha^2}\bigr)\tau}$, so for the first term to be dominating in the right-hand-side of \eqref{KappaNegative_DynamicalSolution} we have to take $\alpha > \lambda$. From this we can say two things. First note that the proper time $\tau$ can take arbitrarily large values, meaning that timelike geodesics are complete when $x^+$ goes to infinity, and then taking the limit $\tau \rightarrow +\infty$ we see that $(x^+,\tilde{x^-}) \rightarrow (+\infty, 0)$, which corresponds exactly to the coordinates of the point $\mathcal{C}$. This means that, at the right-future null infinity, all timelike geodesics go to this point, although they need an infinite amount of proper time to reach it.

We can study the limit of the scalar curvature along one of these timelike geodesics and it turns out that it goes to zero, so an observer will not observer any curvature discontinuity at infinity.

\begin{figure}[hbt]
 \centering
 \includegraphics[scale=1]{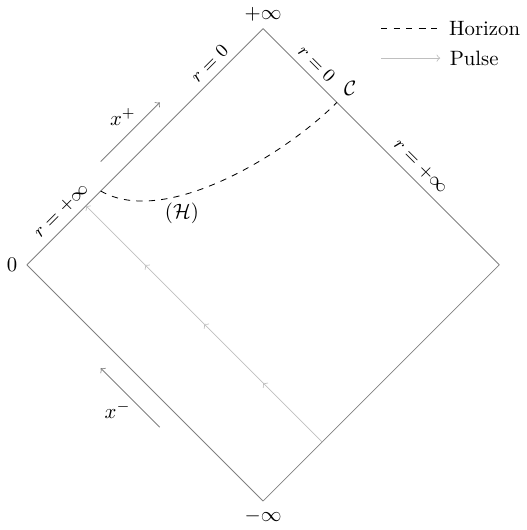}
 \caption{Dynamical spacetime for the Hybrid RST model with $\kappa < 0$. There is no singularity but an apparent horizon appears. The coordinate $r = e^{-\phi}$ is the usual radius.}
 \label{KappaNegative_DynamicSpacetime}
\end{figure}

Note that the spacetime is still singularity free, as in the static case. It is represented on figure \ref{KappaNegative_DynamicSpacetime}.

\section{The hybrid Unruh state}
\label{sec:hybridunruh}

\subsection{Definition}
In section \ref{SubsectionQuantumStates} we defined the Hartle-Hawking and Boulware state by imposing energy conditions at infinity. In the Hartle-Hawking state there is thermal radiation at both past and future null infinity, while particles in the Boulware state do not emit any radiation at infinity. Now in order to get a somewhat more realistic picture of a black hole formation and evaporation we would like to have no radiation at past infinity and thermal radiation of physical particles at future infinity. This situation defines the so-called Unruh state. In appropriate asymptotic coordinates $(\sigma^+(x^+),\sigma^-(x^-))$ these energy conditions translate into
\begin{equation}
 \tilde{T}^{(12)}_{++} = \biggl(\frac{\d x^+}{\d \sigma_+}\biggr) T^{(12)}_{++} \xrightarrow[x^-\rightarrow -\infty]{} 0
 \,,
 \label{EQ_Energy_Condition_Past_Infinity}
\end{equation}
at past infinity and
\begin{equation}
 \tilde{T}^{(12)}_{--} = \biggl(\frac{\d x^-}{\d \sigma_-}\biggr) T^{(12)}_{--} \xrightarrow[x^+\rightarrow +\infty]{} \frac{\lambda^2\kappa_1}{2\pi}
 \,,
 \label{EQ_Energy_Condition_Future_Infinity}
\end{equation}
at future infinity. Note that only the physical particles are concerned by the Unruh state, as the non-physical ones should not emit radiation at past nor future infinity they are still in the Boulware state. Also the Unruh state inevitably leads to a non-static solution as it breaks the $x^+ \leftrightarrow x^-$ symmetry.

\subsection{Boundary conditions}
In order to get a solution which satisfies \eqref{EQ_Energy_Condition_Past_Infinity} and \eqref{EQ_Energy_Condition_Future_Infinity} we have to modify the boundary conditions of physical particles, that is to say the functions $t_{1\pm}(x^\pm)$ defined by \eqref{EQ_Definition_t12} and which appear in the stress-energy tensor $T^{(12)}$ defined by \eqref{RST_T12Definition}. Let us recall that the classical static solution is given by
\begin{equation}
 e^{-2\phi} = -\lambda^2x^+x^- + \frac{M}{\lambda}
 \,.
 \label{EQ_Classical_Static_Solution}
\end{equation}
At null infinity ($x^\pm \rightarrow \pm \infty$) the metric is asymptotically flat as is seen by passing to the coordinates $\sigma^\pm =\pm\frac{1}{\lambda}\ln(\pm\lambda x^\pm)$ in which the metric takes the form $\d s^2 \sim -\d\sigma^+\d\sigma^-$. In these coordinates the energy fluxes are
\begin{equation}
 \tilde{T}^{(12)}_{\pm\pm} = \lambda^2(x^\pm)^2 T^{(12)}_{\pm\pm} = \frac{\lambda^2(x^\pm)^2}{\pi}\sum_{i=1}^2 \kappa_i \bigl(\partial_\pm^2\phi - 2t_{i\pm}\bigr)
 \,.
\end{equation}
From \eqref{EQ_Classical_Static_Solution} we can easily compute the asymptotic behavior of $\partial_\pm^2\phi$ to get
\begin{equation}
 \left\{
 \begin{aligned}
  & \partial_+^2\phi \underset{x^-\rightarrow -\infty}{\sim} \frac{1}{2(x^+)^2}
  \,, \\
  & \partial_-^2\phi \underset{x^+\rightarrow +\infty}{\sim} \frac{1}{2(x^-)^2}
  \,,
 \end{aligned}
 \right.
\end{equation}
so that the simplest choice for the functions $t_{i\pm}$ to satisfy the asymptotic energy conditions \eqref{EQ_Energy_Condition_Past_Infinity} and \eqref{EQ_Energy_Condition_Future_Infinity} is, for physical particles
\begin{equation}
 t_{1+}(x^+) = \frac{1}{4(x^+)^2}
 \,, \quad
 t_{1-}(x^-) = 0
 \,,
 \label{EQ_Unruh_t1}
\end{equation}
and for non-physical particles
\begin{equation}
 t_{2\pm}(x^\pm) = \frac{1}{4(x^\pm)^2}
 \,.
 \label{EQ_Unruh_t2}
\end{equation}
Note that this last equation simply corresponds to the Boulware state. Of course we will have to check once we have the complete solution that the energy conditions \eqref{EQ_Energy_Condition_Past_Infinity} and \eqref{EQ_Energy_Condition_Future_Infinity} are indeed satisfied. The energy fluxes then take the form
\begin{equation}
 \left\{
 \begin{aligned}
  & T^{(12)}_{++} = \frac{\kappa}{\pi}\biggl(\partial_+^2\phi - \frac{1}{2(x^+)^2}\biggr) \\
  & T^{(12)}_{--} = \frac{\kappa}{\pi}\biggl(\partial_-^2\phi - \frac{\kappa_2}{2\kappa (x^-)^2}\biggr)
  \,.
 \end{aligned}
 \right.
\end{equation}
Using null static coordinates $(u,v)$ (see appendix \ref{AppendixCoordinateTransition}) one can check that, on the classical black hole background, this leads to an energy flux at the future horizon $T^{(12)}_{vv} = -\frac{\lambda^2\kappa}{2\pi}$ and at future null infinity $T^{(12)}_{uu} = \frac{\lambda^2\kappa_1}{2\pi}$. This is in agreement with the Unruh state mentioned in \cite{boos_ghost-free_2019}, with the addition here of non-physical particles.

\subsection{Solution}
Having defined the Unruh state using the classical black hole background, we can now study how the back-reaction modifies the geometry of the spacetime. In the absence of matter ($f=0$) the choice \eqref{EQ_Unruh_t1} and \eqref{EQ_Unruh_t2}  for $t_{i\pm}$ leads to
\begin{equation}
 \left\{
 \begin{aligned}
  & u_+(x^+) = -\frac{\kappa}{2}\ln(\lambda x^+) + \alpha_+ x^+ + \beta_+
  \,, \\
  & u_-(x^-) = -\frac{\kappa_2}{2}\ln(-\lambda x^-) + \alpha_- x^- + \beta_-
  \,,
 \end{aligned}
 \right.
\end{equation}
where $\alpha_\pm$ and $\beta_\pm$ are integration constants. The master equation becomes
\begin{equation}
 \Omega = \kappa\phi + e^{-2\phi} = -\lambda^2x^+x^- - \frac{\kappa_2}{2}\ln(-\lambda^2x^+x^-) - \frac{\kappa_1}{2}\ln(\lambda x^+) + \alpha_+ x^+ + \alpha_- x^- + \frac{M}{\lambda}
 \,, 
\end{equation}
where we have set $\beta_+ + \beta_- \equiv \frac{M}{\lambda}$. Note that because of the logarithmic terms we cannot freely translate $x^\pm$ without changing the solution, so the constants $\alpha_\pm$ cannot be set to zero arbitrarily. However, as we have seen previously, adding a linear in $x^\pm$ to the solution corresponds to a perturbation caused by a pulse of classical matter, so the solution for a non-perturbed spacetime is given by
\begin{equation}
 \Omega = \kappa\phi + e^{-2\phi} = -\lambda^2x^+x^- - \frac{\kappa_2}{2}\ln(-\lambda^2x^+x^-) - \frac{\kappa_1}{2}\ln(\lambda x^+) + \frac{M}{\lambda}
 \,.
 \label{EQ_UnruhSolution_Unperturbed}
\end{equation}
When perturbed by a pulse of classical matter along a null geodesic $x^+=x^+_0$ with energy
\begin{equation}
 \frac{1}{2}(\partial_+f)^2 = \frac{m}{\lambda x^+_0}\delta(x^+-x^+_0)
 \,,
\end{equation}
the solution after the pulse ($x^+ > x^+_0)$ becomes
\begin{equation}
 \Omega = \kappa\phi + e^{-2\phi} = -\lambda^2x^+\tilde{x}^- - \frac{\kappa_2}{2}\ln(-\lambda^2x^+x^-) - \frac{\kappa_1}{2}\ln(\lambda x^+) + \frac{M+m}{\lambda}
 \,,
 \label{EQ_UnruhSolution_Perturbed}
\end{equation}
where we use the notation $\tilde{x}^- = x^- + \frac{m}{\lambda^3 x^+_0}$.

In the hybrid Hartle-Hawking/Boulware state discussed previously, the most interesting situation was when the total central charge $\kappa = \kappa_1 + \kappa_2$ was negative. As such we will only discuss the Unruh state for $\kappa < 0$. Note that what we call Unruh state here is really a hybrid quantum state as only the physical particles are in the Unruh state, while the non-physical particles are still in the Boulware state.

\subsection{Unperturbed  hybrid Unruh state solution}
The unperturbed solution for the Unruh state is therefore given by \eqref{EQ_UnruhSolution_Unperturbed}.

\paragraph{Horizon:} we have
\begin{equation}
 \left\{
 \begin{aligned}
  & \partial_+\Omega = 0 \Leftrightarrow  x^+x^- = -\frac{\kappa}{2\lambda^2} > 0
  \,, \\
  & \partial_-\Omega = 0 \Leftrightarrow x^+x^- = -\frac{\kappa_2}{2\lambda^2} > 0
  \,,
 \end{aligned}
 \right.
\end{equation}
so there is no horizon in this spacetime.

\paragraph{Asymptotic behavior:} let us recall the expression for the scalar curvature
\begin{equation}
 R = \frac{8e^{-2\phi}}{\Omega'}\biggl(\partial_+\partial_-\Omega - \frac{\Omega''}{\Omega'^2}\partial_+\Omega\partial_-\Omega \biggr)
 \,,
\end{equation}
with $\partial_+\partial_-\Omega = -\lambda^2$. Since $\Omega' = \kappa - 2e^{-2\phi} < 0$ does not vanish the only possible curvature singularities are on the border of the region \region so we need to study the asymptotic behavior of the curvature on each component of this border.
\begin{enumerate}
 \item \underline{Right future null infinity ($x^+ \rightarrow +\infty)$}: here careful analysis shows that the curvature $R$ goes to zero. This can also be proved by noting that $e^{-2\phi} \sim -\lambda^2x^+x^-$, which corresponds to a flat metric.
 \item \underline{Right past null infinity ($x^- \rightarrow -\infty)$}: again $e^{-2\phi} \sim -\lambda^2x^+x^-$ and $R \rightarrow 0$.
 \item \underline{Left past null infinity ($x^+ \rightarrow 0)$}: analysis shows that
 \begin{equation}
  e^{-2\phi} \sim \alpha^2 (\lambda x^+)(-\lambda x^-)^{\frac{\kappa_2}{\kappa}}
  \,, \quad \alpha \equiv e^{-\frac{M}{\lambda\kappa}} > 0
  \,,
  \label{EQ_AsympMetric_LeftPast}
 \end{equation}
 and note that $\frac{\kappa_2}{\kappa} > 1$. Let us define the asymptotic coordinates $\sigma^\pm(x^\pm)$ by
 \begin{equation}
  \left\{
  \begin{aligned}
   & \lambda \sigma^+ = \frac{1}{\alpha}\ln(\lambda x^+) \\
   & \lambda \sigma^- = \frac{1}{\alpha \Bigl(\frac{\kappa_2}{\kappa}-1\Bigr)(-\lambda x^-)^{\frac{\kappa_2}{\kappa}-1}}
   \,,
  \end{aligned}
  \right.
  \label{EQ_AsympCoord_LeftPast}
 \end{equation}
 the metric becomes asymptotically flat, $\d s^2 \sim -\d\sigma^+\d\sigma^-$, at $x^+ \rightarrow 0$. One can check by a direct computation that $R \rightarrow 0$ in this limit.
 \item \underline{Left future null infinity ($x^- \rightarrow 0)$}: here the behavior of $e^{-2\phi}$ is the same as in the previous case, namely $x^+ \rightarrow 0$, meaning that the metric is asymptotically flat there with the same asymptotic coordinates.
\end{enumerate}
To summarize, this spacetime is singularity and horizon free, and is asymptotically flat.

\subsection{Perturbed hybrid Unruh state solution}
Let us now study what happens when the solution for the Unruh state is perturbed by a pulse of classical matter sent at $x^+ = x^+_0$. Before this shock wave the solution is the same as in the previous section, and after it is given by \eqref{EQ_UnruhSolution_Perturbed}.
\paragraph{Horizon:} there is no horizon before the shock wave, as seen for the unperturbed solution, and after the pulse we have
\begin{equation}
 \left\{
 \begin{aligned}
  & \partial_+\Omega = 0 \Leftrightarrow  x^+\tilde{x}^- = -\frac{\kappa}{2\lambda^2} > 0 \quad (\mathcal{H})
  \,, \\
  & \partial_-\Omega = 0 \Leftrightarrow x^+x^- = -\frac{\kappa_2}{2\lambda^2} > 0
  \,.
 \end{aligned}
 \right.
\end{equation}
The curve $(\mathcal{H})$ is partially contained in the region \region and it corresponds to an apparent horizon, as $(\nabla\phi)^2$ changes sign when crossing it. The asymptote of this horizon when $x^+ \rightarrow +\infty$ is $x^- = x^-_h \equiv -\frac{m}{\lambda^3x^+_0}$.
\paragraph{Asymptotic behavior:} as in the previous case let us study the asymptotic behavior of the metric and curvature. First note that before the shock the spacetime is identical to the unperturbed solution so it asymptotically flat. Let us study the other infinities (for $x^+ > x^+_0$) now.
\begin{enumerate}
 \item \underline{Right future null infinity $(x^+ \rightarrow +\infty$)}
 \begin{enumerate}
  \item before the horizon ($\tilde{x}^-  < 0$): in this case $e^{-2\phi} \sim -\lambda^2 x^+\tilde{x}^-$ so the metric is asymptotically flat.
  \item behind the horizon ($\tilde{x}^-  > 0$): this limit is a bit more involved as
 \begin{equation}
  e^{-2\phi} \sim \alpha^2 e^{2\frac{\lambda^2}{\kappa}x^+\tilde{x}^-} (\lambda x^+)(-\lambda x^-)^{\kappa_2/\kappa}
  \,, \quad
  \alpha = e^{-\frac{M+m}{\lambda\kappa}}
  \,,
 \end{equation}
 and it is not obvious how to find asymptotically flat coordinates. However a direct computation shows that
 \begin{equation}
  R \sim -\frac{8\lambda^2}{\kappa}e^{-2\phi} \rightarrow 0
  \,,
 \end{equation}
 so spacetime is also flat there.
 \end{enumerate}  
 \item \underline{Right past null infinity ($x^- \rightarrow -\infty$)}: there $e^{-2\phi} \sim -\lambda^2x^+x^-$ so actually nothing changes compared to the unperturbed solution, the metric is still flat.
 \item \underline{Left past null infinity ($x^+\rightarrow 0$)}: since this border is located before the shock the perturbation does not change anything and spacetime is also flat.
 \item \underline{Left future null infinity ($x^- \rightarrow 0$)}: here the situation is almost the same as for the unperturbed case, as
 \begin{equation}
  e^{-2\phi} \sim \alpha^2 e^{-2\frac{\lambda^2}{\kappa}x^+x^-_h} (\lambda x^+)(-\lambda x^-)^{\frac{\kappa_2}{\kappa}}
  \,, \quad \alpha \equiv e^{-(M+m)/\lambda\kappa}
  \,.
  \label{EQ_AsymMetric_LeftFuture_Perturbed}
 \end{equation}
This is similar to the unperturbed situation, and one can check that the curvature indeed goes to zero.
\end{enumerate}
As far as spacetime geometry is concerned we can see that the hybrid Unruh state gives a very similar solution to the one studied previously for the hybrid Hartle-Hawking/Boulware state. The only notable difference is that the unperturbed solution is no longer static, because of the inherent nature of the Unruh state. But of course the main motivation behind the Unruh state is to determine whether the energy fluxes are consistent with what we wanted to observe, namely radiation at future infinity without radiation at past infinity. This is what we will discuss now.

\section{Energy fluxes}\label{sec:energy}

In this section we will compute the energy fluxes on the spacetime border of the hybrid Hartle-Hawking/Boulware and the hybrid Unruh states for both the unperturbed and perturbed cases. To reduce the amount of computation one can study these fluxes in the perturbed case, and then take the limit $m \rightarrow 0$ to get the corresponding flux in the unperturbed situation. Note that the apparent horizon disappears in this limit. On each border component we are interested in computing the outgoing or ingoing flux $\tilde{T}^{(12)}_{\pm\pm}$ in asymptotically flat coordinates.

\subsection{Energy for the hybrid Hartle-Hawking/Boulware state}
Let us go back to the hybrid Hartle-Hawking/Boulware state discussed earlier and compute the outgoing or ingoing energy flux at each border.
\begin{enumerate}
 \item \underline{Right future null infinity $(x^+ \rightarrow +\infty$)}
 \begin{enumerate}
  \item before the horizon ($\tilde{x}^-  < 0$): here $e^{-2\phi} \sim -\lambda^2x^+\tilde{x}^-$ so we have the asymptotically flat coordinates $\sigma_+ = \frac{1}{\lambda}e^{\lambda x^+}$ and $\sigma^- = -\frac{1}{\lambda}e^{-\lambda\tilde{x}^-}$. The outgoing energy flux is then
  \begin{equation}
   \tilde{T}^{(12)}_{--} = \frac{1}{\pi}\biggl(\frac{\d x^-}{\d \sigma^-}\biggr)^2 \biggl(\partial_-^2\phi - \frac{\kappa_2}{2(x^-)^2}\biggr) \rightarrow 
   \frac{\lambda^2}{2\pi}\biggl(\kappa - \frac{(\tilde{x}^-)^2}{(x^-)^2}\kappa_2\biggr)
   \,.
   \label{fluxHH}
  \end{equation}
  We discussed this result in \cite{potaux_space-time_2022}, arguing that one could define a radiation entropy and recover a Page Curve, meaning the whole information is recovered.
  
  Taking the limit $m \rightarrow 0$, \textit{i.e.} $\tilde{x}^- = x^-$, we obtain that in the unperturbed solution without horizon there is still radiation as $T^{(12)}_{--} \rightarrow \frac{\lambda^2\kappa_1}{2\pi}$ for all values of $x^-$.
  
  \item behind the horizon ($\tilde{x}^-  > 0$): since
  \begin{equation}
   e^{-2\phi} \sim \alpha^2e^{\frac{2\lambda^2}{\kappa}x^+\tilde{x}^-}(-\lambda^2x^+x^-)^{\kappa_2/\kappa}
   \,,
  \end{equation}
  the asymptotic coordinates on this border are far from obvious. However it is rather straightforward to show that $T^{(12)}_{--} \rightarrow 0$ there, so we can consider that there is no outgoing flux behind the horizon.
 \end{enumerate}
 \item \underline{Right past null infinity ($x^- \rightarrow -\infty$)}: with and without perturbation we have $e^{-2\phi} \sim -\lambda^2x^+x^-$ so we can define the asymptotically flat coordinates $\sigma^\pm = \pm\frac{1}{\lambda}e^{\pm\lambda x^\pm}$ and get that the incoming energy flux is
 \begin{equation}
  \tilde{T}^{(12)}_{++} = \frac{1}{\pi}\biggl(\frac{\d x^+}{\d \sigma^+}\biggr)^2 \biggl(\partial_+^2\phi - \frac{\kappa_2}{2(x^+)^2}\biggr) \rightarrow 
   \frac{\lambda^2\kappa_1}{2\pi}
   \,.
 \end{equation}
 This energy flux does not appear to have a physical reason to exist, hence the need to define the Unruh state where it is absent.
 \item \underline{Left past null infinity ($x^+ \rightarrow 0$)}: as this is before the pulse it has no influence. We have $e^{-2\phi} \sim \alpha^2(-\lambda^2x^+x^-)^{\kappa_2/\kappa}$ so we can find asymptotic coordinates such that
 \begin{equation}
  \frac{d \sigma^\pm}{\d x^\pm} = \frac{1}{\alpha(\pm \lambda x^\pm)^{\kappa_2/\kappa}}
  \,,
 \end{equation}
 and obtain that $\tilde{T}^{(12)}_{--} \rightarrow 0$, meaning that there is no incoming energy.
 \item \underline{Left future null infinity ($x^- \rightarrow 0$)}: a similar reasoning to the previous limit leads to $\tilde{T}^{(12)}_{++}$, so there is no outgoing radiation there.
\end{enumerate}

\begin{figure}[hbt]
 \centering
 \includegraphics{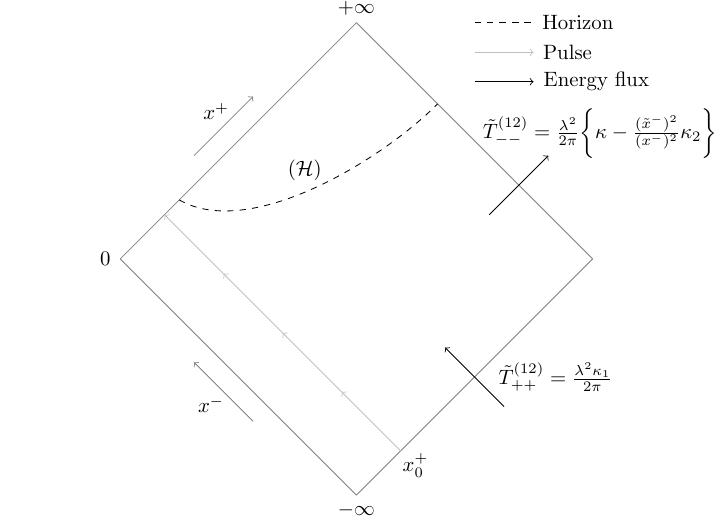}
 \caption{Spacetime for the hybrid Hartle-Hawking/Boulware state perturbed by a pulse of classical matter. An apparent horizon is created there is radiation at both past and future infinity. Note that without the pulse the horizon disappears but there is still thermal radiation.}
 \label{FIG_KappaNegative_Energy}
\end{figure}

\subsection{Energy for the hybrid Unruh state}
\begin{figure}[hbt]
 \centering
 \includegraphics{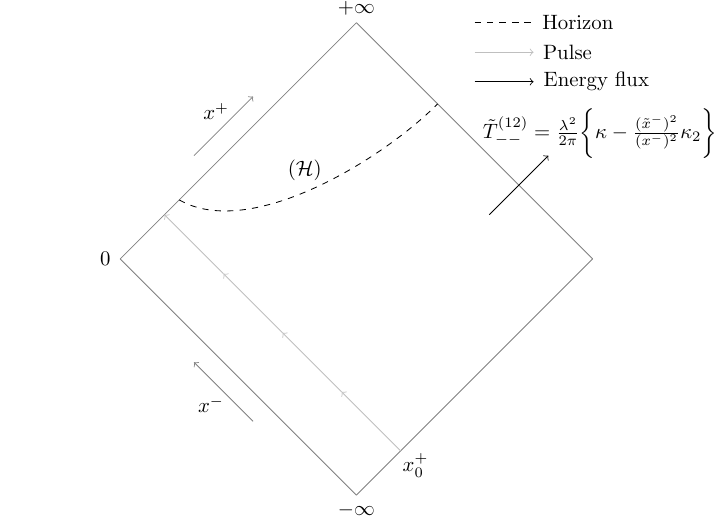}
 \caption{Spacetime for the hybrid Unruh/Boulware state perturbed by a pulse of classical matter. An apparent horizon is created and there is Hawking radiation at future infinity with modifications at later retarded times.}
 \label{FIG_UnruhPerturbed}
\end{figure}

We can follow the same procedure for the Unruh state and the results will be the same except for one important difference which is that there is no incoming radiation at the right past infinity, \textit{i.e.}
\begin{equation}
 \tilde{T}^{(12)}_{++} \rightarrow 0
 \,,
\end{equation}
when $x^- \rightarrow -\infty$. This is in accordance with the definition of the Unruh state. Note that on the right future infinity, before the horizon, we still have
\begin{equation}
 \tilde{T}^{(12)}_{--}\rightarrow \frac{\lambda^2}{2\pi}\biggl(\kappa - \frac{(\tilde{x}^-)^2}{(x^-)^2}\kappa_2\biggr)
 \,,
 \label{fluxU}
\end{equation}
so once again, as done in \cite{potaux_space-time_2022} for the Hartle-Hawking/Boulware state, one can define an entropy radiation and show that it exhibits the Page Curve. Therefore with the Unruh state we have a completely regular solution with an apparent horizon and where the only energy flux is an outgoing one corresponding to Hawking radiation. This is presented on figure \ref{FIG_UnruhPerturbed}.

We also note that the spacetime corresponding to a Unruh state is non-static so that the energy is not conserved. This is seen in our energy analysis: zero incoming energy
at past null infinity is transformed to  a non-vanishing flux of energy at the right future infinity. This flux is thermal for earlier retarded times while it gets the important corrections
as in (\ref{fluxU}) for later times. These corrections are responsible for the restoration of the initially seemingly lost information that is manifested in the declining
branch in the Page curve for the entropy of the asymptotic radiation.

\subsection{Entropy of  asymptotic radiation} 
Here we present a discussion of the entropy of the asymptotic radiation as is seen at the right future infinity.  We notice that in the context of the Page curve one usually computes the entanglement entropy.  In the RST model it is given by value of the field $\psi$ at the horizon, see  \cite{Myers:1994sg,Solodukhin:1994yz}. 
Instead, we present here analysis of the thermal entropy in asymptotic infinity. This analysis is common both for the hybrid Hartle-Hawking and the Unruh quantum states.

In \cite{potaux_space-time_2022} we studied the asymptotic energy radiation at the right future null infinity and we suggested that the radiation entropy could be defined by the equation
\begin{equation}
 \partial_- S = 2\pi(-x^-)T^{(12)}_{--}
 \,,
 \label{entropy}
\end{equation}
where $T^{(12)}_{--}$, defined by \eqref{RST_T12Definition}, is expressed here as
\begin{equation}
 \pi T^{(12)}_{--} = \kappa\partial_-^2\phi - \frac{\kappa_2}{2(x^-)^2}
 \,.
\end{equation}
As is seen from the asymptotic values of the energy density (\ref{fluxHH}) and (\ref{fluxU}) the energy flux is being positive for earlier values of $x^-$ changes sign for $x^-> x^-_m$,
$x_m=(1-\sqrt{\frac{\kappa}{\kappa_2}})^{-1}x^-_h$.
So that the entropy defined by (\ref{entropy}) while growing for $x^-<x^-_h$ becomes decreasing for times $x^-> x^-_m$. This is precisely the behavior
expected for the Page curve in an information preserving scenario.

We can compute the variation of the radiation entropy along the right future null infinity by computing, for a finite value of $x^+$,
\begin{equation}
 \Delta S = 2\pi\int_{-L}^{-\varepsilon}\d x^-\,(-x^-)T^{(12)}_{--}
 \,,
\end{equation}
where $\varepsilon,L > 0$ are two regulators designed to regularize the possible divergences at $x^-\rightarrow0$ and $x^- \rightarrow -\infty$ respectively. Then we will take the limits $\varepsilon \rightarrow 0$ and $L \rightarrow +\infty$ and finally $x^+ \rightarrow +\infty$. Note that we can assume $-\varepsilon > x^-_h$ and $-L < x^-_h$. The result of this computation was presented in \cite{potaux_space-time_2022} and we will present the important steps here. Replacing $T^{(12)}_{--}$ by its expression and performing an integration by parts leads to
\begin{equation}
 \Delta S = \Bigl[\kappa_2\ln(-x^-) + 2\kappa(\phi-x^-\partial_-\phi)\Bigr]_{x^-=-L}^{x^-=-\varepsilon}
 \equiv \Bigl[s(x^-)\Bigr]_{x^-=-L}^{x^-=-\varepsilon}
 \,,
\end{equation}
so we need to compute the behavior of the quantity $s(x^-)$ for $x^-=-\varepsilon \rightarrow 0$ and $x^-=-L \rightarrow -\infty$. Note that for the moment we work with a finite value of $x^+$.

\bigskip

\noindent{\bf 1.}  \ For $x^-=-\varepsilon \rightarrow 0$ we have (we remind that $\kappa=\kappa_1+\kappa_2<0$ and $\kappa_2<0$)
 \begin{equation}
  \Omega = \kappa\phi + e^{-2\phi} = \lambda^2x^+\varepsilon - \frac{\kappa_2}{2}\ln\varepsilon + C
  \,,
  \label{Ophi}
 \end{equation}
 where $C$ is a constant, which depends on $x^+$, given by
 \begin{equation}
  C = \lambda^2x^+x^-_h - \frac{\kappa_2}{2}\ln(\lambda^2x^+) + \frac{M+m}{\lambda}
  \,.
  \label{RST_DeltaS_CDefinition}
 \end{equation}  
 Therefore  $\Omega \rightarrow -\infty$ when $\varepsilon \rightarrow 0$ and hence one has that $\phi \rightarrow +\infty$. Thus
 \begin{equation}
  \kappa\phi = -\frac{\kappa_2}{2}\ln\varepsilon + C + O(\epsilon)
  \,.
 \end{equation}
 Then differentiating (\ref{Ophi})  it is straightforward to show that
 \begin{equation}
  \kappa\varepsilon\partial_-\phi \rightarrow \frac{\kappa_2}{2}
  \,,
 \end{equation}
when $x^-=-\epsilon\rightarrow 0$, which leads to
 \begin{equation}
  s(-\varepsilon) = 2C + \kappa_2\, , \ \  \varepsilon\rightarrow 0
  \,.
 \end{equation}
 
 \medskip
 
 \noindent{\bf 2.}
 Let us now consider the limit $x^- = -L \rightarrow -\infty$. We have
 \begin{equation}
  \Omega = \kappa\phi + e^{-2\phi} = \lambda^2x^+L - \frac{\kappa_2}{2}\ln L + C
  \,,
  \label{RST_DeltaS_Omega_L}
 \end{equation}
 where $C$ is the same constant as in \eqref{RST_DeltaS_CDefinition}. Since $\Omega \sim \lambda^2x^+L \rightarrow +\infty$ we have that $\phi \rightarrow -\infty$ with $e^{-2\phi} \sim \lambda^2x^+L$, meaning that $\phi \sim -\frac{1}{2}\ln L$ for large $L$.
 Inserting this into \eqref{RST_DeltaS_Omega_L} gives
 \begin{equation}
  e^{-2\phi} = \lambda^2x^+L + \frac{\kappa_1}{2}\ln L + C +O(1/L)
  \,.
 \end{equation}
 Then it is once again easy to show that
 \begin{equation}
  L\partial_-\phi \rightarrow \frac{1}{2}
  \,,
 \end{equation}
 in the limit $-x^-=L\rightarrow \infty$,
 which leads to
 \begin{equation}
  s(-L) = -\kappa_1\ln L + \kappa(1-\ln(\lambda^2x^+)) 
  \,.
 \end{equation}
 in this limit.
Now we can write the change in entropy between $x^- = -L$ and $x^-=-\varepsilon$ for a fixed value of $x^+$ as
\begin{equation}
 \Delta S = s(-\varepsilon) - s(-L) = \kappa_1\ln(\lambda^2x^+L/e)  + \frac{2M}{\lambda} + \frac{2m}{\lambda}\biggl(1-\frac{x^+}{x^+_0}\biggr)
 \,,
 \label{DS}
\end{equation}
 using the expression \eqref{RST_DeltaS_CDefinition} for the constant $C$ as well as that $x^-_h = -\frac{m}{\lambda^3x^+_0}$. This is indeed the result presented in 
 \cite{potaux_space-time_2022}.
As was discussed in   \cite{potaux_space-time_2022} the first term in the above equation represents the entropy of thermal radiation (at the Hawking temperature $T_H=\lambda/2\pi$)
that is seen by passing to the 
asymptotic coordinates $(\sigma^+\,, \sigma^-)$. The second term represents the entropy of the classical black hole of mass $M$ while the third term is decreasing in $x^+$ and it represents the contribution due to the classical matter perturbation. The other observation is that the sum of two last terms in (\ref{DS}) is in fact the classical Wald's entropy 
$S_{W}=2e^{-2\phi}$ computed for the perturbed classical solution  (\ref{dilatonPhi}) at the unperturbed horizon $x^-=0$.  This is  so at least for relatively small values of $x^+$.
The significance of this observation is not clear at the present point and  perhaps deserves the further study.
We also note that  the change in the entropy (\ref{DS}) does not depend on the parameter $\kappa_2$ due to the non-physical particles.

\section{Concluding remarks}\label{sec:conclude}

\begin{table}[hbt]
\begin{tabular}{|c||c|c|c|c|c|c|}
\hline
State
& HH
& B (non-physical)
& B (physical)
& Hy ($\kappa =0)$
& Hy ($\kappa >0)$
& Hy ($\kappa <0)$
\\
\hline
$(\kappa_1, \kappa_2)$
& $(>0,=0)$
& $(<0,=0)$
& $(>0,=0)$
& $(>0,<0)$
& $(>0,<0)$
& $(>0,<0)$
\\
\hline
Horizon
& Killing
& None
& None
& None
& None
& None
\\
\hline
Singularity
& Spacelike
& None
& Null
& Timelike
& Timelike
& None
\\
\hline
Geodesics
& Incomplete
& Complete
& Incomplete
& Incomplete
& Incomplete
& Complete
\\
\hline
WH Type I
& No
& No
& Yes
& No
& No
& No
\\
\hline
WH Type II
& No
& No
& No
& No
& No
& Yes
\\
\hline
\end{tabular}
\caption{Main properties of the static solutions of each case studied. HH stands for Hartle-Hawking, B for Boulware and Hy for Hybrid. WH stands for wormhole.}
\label{Table_Recap}
\end{table}
We have summarized the main properties of every static solution to the RST model studied in this paper in table \ref{Table_Recap}. When a singularity and (or) a horizon are present, sending in a shock wave of classical matter to get a dynamical spacetime modifies their trajectory, while for solutions containing neither of these two things the shock has the effect of creating a horizon.
Our analysis reveals that as soon as at least one (physical or non-physical) particle in the multiplet of fields is in the Boulware quantum state, the horizon disappears in the
back-reacted geometry. Instead, the norm of the Killing vector may have either a minimum that signals a wormhole structure which mimics a black hole horizon, or be monotonically decreasing
with the spacetime ending at either a timelike or lightlike singularity. The global structure of the back-reacted spacetime would be quite radically different from the
classical CGHS black hole. We expect that a similar behavior should be valid in the four-dimensional situation when the back-reaction is taken into account in a self-consistent manner.
The respective analysis is of course much more complicated due to the increasing technical difficulties. Although performing this analysis becomes urgent and practical given the potential possibility
of seeing the deviations from the classical black hole structure in the gravitational wave experiments.

 As explained here and in \cite{potaux_space-time_2022}, the most interesting scenario is the hybrid state when there are more non-physical particles than physical ones ($\kappa < 0$). Note that this solution is similar to the Boulware solution for non-physical particles, the differences being that it is geodesically complete, has the structure of a black hole mimicker suggested in \cite{damour_wormholes_2007}, and contains the thermal radiation due to physical particles (as only physical particles radiate energy at infinity, see \cite{potaux_space-time_2022} for more details about energy radiation). The fact that one can recover the Page curve for the change in entropy of radiation could suggest that this is a possible realization of an object which mimics a black hole without allowing any information loss. An interpretation in terms of creation of physical and non-physical particle pairs was proposed in \cite{potaux_space-time_2022} to try to understand this result conceptually. In the future, it will be interesting to study these aspects more closely and investigate the connection of our work with the recent and on-going studies on quantum extremal surfaces.

\vspace{0.2 cm}

\noindent {\bf Acknowledgements} The work of DS is supported by the DST-FIST grant number SR/FST/PSI-225/2016 and SERB MATRICS grant MTR/2021/000168.

\newpage

\appendix
\section{Coordinate choice}
\subsection{Conformal coordinates}
\label{AppendixConformalCoord}
In a two dimensional spacetime one can always perform a change of coordinates to put the metric under the conformal form (see for instance \cite{dubrovin_modern_1984})
\begin{equation}
 \d s^2 = e^{2\rho}(-\d t^2 + \d x^2)
 = -e^{2\rho}\d x^+ \d x^-
 \,, \ \rho=\rho(x^+, x^-)\, ,
 \label{Appendix_ConformalMetric}
\end{equation}
where the light-cone coordinates $x^\pm$ are defined by
\begin{equation}
 x^\pm = t \pm x
 \,.
\end{equation}
The non-vanishing metric coefficients are therefore
\begin{equation}
 g_{+-} = -\frac{1}{2}e^{2\rho}
 \,, \quad
 g^{+-} = -2e^{-2\rho}
 \,,
\end{equation}
which give the following Christoffel symbols
\begin{equation}
 \tensor{\Gamma}{^+_+_+} = 2\partial_+\rho
 \,, \quad
 \tensor{\Gamma}{^-_-_-} = 2\partial_-\rho
 \,,
\end{equation}
and the Ricci scalar is given by
\begin{equation}
 R = \frac{2}{g}R_{+-+-} = 8e^{-2\rho}\partial_+\partial_-\rho
 \,.
\end{equation}
Note that since we are working in two dimensions, $R_{+-+-}$ is the only independent component of the Riemann tensor and spacetime is flat if and only if $R=0$. For a scalar function $f$ we have
\begin{equation}
 (\nabla f)^2 = -4e^{-2\rho}\partial_+f\partial_-f
 \,,
\end{equation}
and
\begin{equation}
 \square f = -4e^{-2\rho}\partial_+\partial_- f
 \,.
\end{equation}
As explained in details in the main text these conformal coordinates are well suited when treating solutions perturbed by an incoming pulse of energy.
\subsection{Static coordinates}
\label{AppendixCoordinateTransition}
In \cite{potaux_quantum_2022} we had exclusively worked in static coordinates $(t,\phi)$ in which the metric is written as
\begin{equation}
 \d s^2 = -g(\phi)\d t^2 + \frac{h^2(\phi)}{g(\phi)}\d \phi^2
 \,,
 \label{Appendix_StaticMetric}
\end{equation}
where the coordinate $\phi$ corresponds to the value of the dilaton field $\phi$. By definition these coordinates are only suited to describe static solutions and are therefore not adapted to discuss the dynamical situations considered in this paper. However, for static solutions, the study of geodesic completeness is rather easy with this metric, as we will see in the following.

For now let us discuss how to transition from the conformal coordinates to the static ones for the solution \eqref{RST_OmegaStaticSolutionKappa2=0}. In \cite{potaux_quantum_2022} we found that the functions $g(\phi)$ and $h(\phi)$, which determine the static metric \eqref{Appendix_StaticMetric}, could be written as 
\begin{equation}
 g(\phi) = e^{2\phi}Z(\phi)
 \,, \quad
 h(\phi) = \frac{1}{2\lambda}e^{2\phi}Z'(\phi)
 \,,
\end{equation}
with the function $Z(\phi)$ determined by the equation
\begin{equation}
 \Omega(\phi) = \kappa\phi + e^{-2\phi} = Z(\phi) + A\ln Z(\phi) + a
 \,, \quad
 a \in \mathbb{R}
 \,,
\end{equation}
where
\begin{equation}
 A = \frac{\kappa}{8\lambda^2}C(C+4\lambda)
 \,, \quad C \in \mathbb{R}
 \,.
\end{equation}
Note that this solution, and the following discussion, is easily extended to the hybrid case where there are two constants $(A_1, A_2)$ for each of the two types of particles considered.

Let us define
\begin{equation}
 y(\phi) \equiv \frac{1}{2\lambda}\ln Z(\phi)
 \,,
\end{equation}
such that the static metric takes the form
\begin{equation}
 \d s^2 = e^{2\phi}e^{2\lambda y}(-\d t^2 + \d y^2)
 \,,
\end{equation}
and
\begin{equation}
 \Omega = \kappa\phi + e^{-2\phi} = e^{2\lambda y} + 2\lambda A y + a
 \,.
\end{equation}
Then we define the null coordinates
\begin{equation}
 \begin{aligned}
  \sigma^+ & \equiv t + y
  \,, \\
  \sigma^- & \equiv t - y
  \,, \\
 \end{aligned}
\end{equation}
and we get
\begin{equation}
 \d s^2 = -e^{2\phi}e^{\lambda(\sigma^+-\sigma^-)}\d\sigma^+\,\d\sigma^-
 \,.
\end{equation}
Finally let us define
\begin{equation}
 x^\pm \equiv \pm \frac{1}{\lambda}e^{\pm\lambda \sigma^+}
 \,,
\end{equation}
so that the metric becomes
\begin{equation}
 \d s^2 = -e^{2\phi}\d x^+ \d x^-
 \,,
\end{equation}
and
\begin{equation}
 \Omega = \kappa\phi + e^{-2\phi} = -\lambda^2x^+x^- + A\ln(-\lambda^2x^+x^-) + a
 \,,
\end{equation}
\begin{equation}
 Z = e^{2\lambda y} = -\lambda^2x^+x^-
 \,.
\end{equation}
This suggests that the constants $P$ and $M$ in \eqref{RST_OmegaStaticSolutionKappa2=0} are related to the constants $A$ and $a$ by
\begin{equation}
 P = \frac{A}{2\kappa} = \frac{C(C+4\lambda)}{16\lambda^2}
 \,, \quad
 a = \frac{M}{\lambda}
 \,.
\end{equation}
In \cite{potaux_quantum_2022} we found that the Hartle-Hawking state corresponds to $A=0$, \textit{i.e.}\ $P=0$, and the Boulware state to $A = - \frac{\kappa}{2}$, \textit{i.e.}\ $P = -\frac{1}{4}$. This is consistent with what we found here in section \ref{SubsectionQuantumStates}.

\section{Geodesics}
\label{AppendixConformalGeodesics}
In this section we will explain how one can study geodesic completeness of the various solutions presented in this paper. In order to do this one needs to determine whether the affine parameter associated to each geodesics can take arbitrary large values, in which case the spacetime is geodesically complete. Otherwise it is incomplete. It turns out that null geodesics are easily studied in the conformal metric \eqref{Appendix_ConformalMetric} while for timelike geodesics the most adapted coordinates are the static ones \eqref{Appendix_StaticMetric}. Note that timelike geodesics are therefore difficult to study in dynamical solutions, where we cannot use static coordinates, and can only be solved pertubatively in some cases.
\subsection{Null geodesics}
Let us use the conformal coordinates $(x^+,x^-)$ where the metric is given by \eqref{Appendix_ConformalMetric}. The geodesic equations are, for an affine parameter $\chi$,
\begin{equation}
 \frac{\d^2x^\pm}{\d\chi^2} + 2\partial_\pm\rho\biggl(\frac{\d x^\pm}{\d\chi}\biggr)^2 = 0
 \,.
\end{equation}
Note that we consider a general situation when spacetime is not necessarily static and generic function $\rho(x^+, x^-)$.
For timelike geodesics these equations are not easy to solve, which is the reason why it is preferable to use static coordinates, as will be explained later on. However null geodesics, which satisfy $\d s^2=-e^{2\rho}\d x^+\d x^-=0$, are curves of constant $x^+$ or constant $x^-$ so that the geodesic equations are greatly simplified and can be solved analytically. Let us consider a null geodesic defined by $x^+ = x^+_0$ and parametrized by an affine parameter $\chi$ (of course the following discussion is easily adapted to a geodesic defined by $x^- = x^-_0$). Then $\frac{\d x^+}{\d\chi} = 0$ along this lightlike geodesic which means that
\begin{equation}
 \frac{\d\rho}{\d\chi} = \frac{\d x^-}{\d\chi}\partial_-\rho
 \,,
\end{equation}
so the geodesic equation for $x^-$ becomes
\begin{equation}
 \biggl(\frac{\d x^-}{\d\chi}\biggr)^{-1}\frac{\d^2x^-}{\d\chi^2}
 = -2\frac{\d\rho}{\d\chi}
 \,,
\end{equation}
and this can be integrated into
\begin{equation}
 \ln\frac{\d x^-}{\d\chi} = -2\rho + c
 \,, \quad
 c \in \mathbb{R}
 \,,
\end{equation}
where we chose $\frac{\d x^-}{\d\chi} > 0$, which corresponds to a future-oriented geodesic. Redefining the affine parameter $\chi$ by $\chi \rightarrow e^c\chi$ this leads to
\begin{equation}
 \d\chi = e^{2\rho}\d x^-
 \,,
\end{equation}
so for a geodesic of constant $x^+=x^+_0$ going from $x^-_0$ to some value of $x^-$ the variation of the affine parameter $\chi$ is given by the integral
\begin{equation}
 \Delta\chi = \int_{x^-_0}^{x^-}\d u\,e^{2\rho(x^+_0 , u)}
 \,.
 \label{DeltaLambdaConformal}
\end{equation}
This geodesic is complete if and only if this integral is divergent for limiting values of $x^-$. Of course there is the analog integral for geodesics of constant $x^-=x^-_0$.

\subsection{Timelike geodesics}
\label{AppendixTimelikeGeodesics}
As mentioned previously it is rather difficult to study timelike geodesics in the conformal metric \eqref{Appendix_ConformalMetric}. Since we have seen how to transition between conformal and static coordinates let us use the static metric given by \eqref{Appendix_StaticMetric}. The non-vanishing Christoffel symbols are
\begin{equation}
 \tensor{\Gamma}{^t_t_\phi} = \frac{g'}{2g}
 \,, \quad
 \tensor{\Gamma}{^\phi_t_t} = \frac{gg'}{2h^2}
 \,, \quad
 \tensor{\Gamma}{^\phi_\phi_\phi} = \frac{h'}{h} - \frac{g'}{2g}
 \,,
\end{equation}
so the geodesic equation for the time coordinate $t$ is
\begin{equation}
 g\frac{\d^2t}{\d\tau^2} + g'\frac{\d\phi}{\d\tau}\frac{\d t}{\d\tau} = \frac{\d}{\d\tau}\biggl(g\frac{\d t}{\d\tau}\biggr) = 0
 \,,
\end{equation}
which means that $g\frac{\d t}{\d\tau} = E$ is a constant of motion for the geodesic. Note that we denote the affine parameter by $\tau$ as it corresponds to the proper time along timelike geodesics. Then we can use the condition $\bigl(\frac{\d s}{\d \tau}\bigr)^2 = -1$ to get
\begin{equation}
 \biggl(\frac{\d\phi}{\d\tau}\biggr)^2 = \frac{E^2-g}{h^2}
 \,. 
\end{equation}
The variation of the proper time $\tau$ is then given by
\begin{equation}
 \Delta\tau = \int_{\phi_0}^\phi \d\tilde{\phi}\,\sqrt{\frac{h^2(\tilde{\phi})}{E^2-g(\tilde{\phi})}}
 \,.
 \label{DeltaLambdaConformalTimelike}
\end{equation}
We can thus study the completeness of timelike geodesics by examining whether this integral is convergent or divergent for limiting values of $\phi$. However note once again that this is only possible for static solutions, as dynamical spacetimes require using conformal coordinates in which timelike geodesics are not easily determined.
\section{Tangent vectors of implicit curves}
\label{Appendix_TimelikeSpacelike}
When discussing the nature of singularities or horizons, that is whether they are timelike, null or spacelike, it is useful to have a concrete criteria. Thus let us consider an implicit curve defined by $F(x^+,x^-) = 0$. A tangent vectors $\zeta$ is
\begin{equation}
 (\zeta^+,\zeta^-) = e^{-\phi}\biggl(1, -\frac{\partial_+F}{\partial_-F}\biggr)
 \,,
\end{equation}
and its norm is
\begin{equation}
 \zeta^2 = \frac{\partial_+F}{\partial_-F}
 \,.
\end{equation}
Therefore the singularity is timelike if $\frac{\partial_+F}{\partial_-F} < 0$ and spacelike if $\frac{\partial_+F}{\partial_-F} > 0$. We use this useful criteria several times in the main text.

\end{document}